 \documentclass[final,5p,times,twocolumn,authoryear]{elsarticle}

\usepackage{amssymb}
\usepackage{lipsum}
\usepackage{amsthm}
\usepackage[colorlinks=true,allcolors=blue]{hyperref}

\usepackage{enumitem}
\setitemize{noitemsep}

\usepackage[labelfont=bf]{caption}

\usepackage[linesnumbered,ruled,vlined]{algorithm2e}
\DontPrintSemicolon

\SetKwComment{Comment}{\color{green!50!black}// }{}

\SetKwProg{Function}{function}{}{}

\usepackage{multirow}
\usepackage[table,xcdraw]{xcolor}

\journal{Astronomy $\&$ Computing}

\begin{document}

\begin{frontmatter}



\title{3D radio data visualisation in open science platforms for next-generation observatories}


\author{I. Labadie-García}
\ead{ixakalab@iaa.es}
\author{J. Garrido}
\author{L. Verdes-Montenegro}
\author{M.Á. Mendoza}
\author{M. Parra-Royón}
\author{S. Sánchez-Expósito}
\author{R. Ianjamasimanana}

\affiliation{organization={Instituto de Astrofísica de Andalucía},%Department and Organization
            addressline={Glorieta de la Astronomía s/n}, 
            city={Granada},
            postcode={18008},
            country={Spain}}
            
%\address{Instituto de Astrofísica de Andalucía, Glorieta de la Astronomía s/n, Granada, 18008, Spain}

\begin{abstract}
Next-generation telescopes will bring groundbreaking discoveries but they will also present new technological challenges. The Square Kilometre Array Observatory (SKAO) will be one of the most demanding scientific infrastructures, with a projected data output of 700 PB per year to be distributed to a network of SKA Regional Centres. Current tools are not fully suited to manage such massive data volumes, therefore, new research is required to transform science archives from data providers into service providers. In this paper we examine how a science archive can deliver advanced visualisation capabilities for the SKA science archive.
In particular, we have conducted a thorough exploration of existing visualisation software for astronomy and other fields to identify tools capable of addressing Big Data requirements. Using selected technologies, we have developed a prototype archive that provides access to interactive visualisations of 3D radio data through web-based interfaces, adhering to International Virtual Observatory Alliance (IVOA) recommendations to favour interoperability and Open Science practices.
In addition, we discuss how current IVOA recommendations support these visualisation capabilities and how they could be expanded. Our prototype archive includes a service to generate 3D models on the fly as a server operation, enabling remote visualisations in a flexible manner; for instance, a set of parameters can be used to customise the models and their visualisation.
We have used SKA precursor and pathfinder data to test its usability and scalability, concluding that remote visualisation is a viable solution for handling high-volume data. However, our prototype is constrained by memory limitations, requiring techniques to reduce memory usage.
\end{abstract}




\end{frontmatter}
\tableofcontents



\section{Introduction}\label{introduction}

The amount of available astrophysical data has been steadily increasing in recent years due to the construction of new telescopes. This trend is expected to continue in the future and will further escalate with the creation of next-generation observatories. Among these, the Square Kilometre Array Observatory\footnote{\url{https://www.skao.int/en}} (SKAO) stands out for the vast quantity and complexity of data it will produce.
SKAO has begun the construction of two radio interferometers at two sites \citep{SKA_ConstructionProporsal}: one in Western Australia (SKA-Low) and another in the Karoo region of South Africa (SKA-Mid). SKA-Low will consist of 131,072 log-periodic antennas divided into 512 stations, with a maximum baseline of 74 km, observing at a frequency range of $50 - 350$ MHz. 
SKA-Mid will consist of 197 steerable dishes, including those of the MeerKAT telescope\footnote{\url{https://www.sarao.ac.za/gallery/meerkat/}} \citep{meerkat2016}, with a maximum baseline of 150 km, observing in a frequency range between 350 MHz and 15.4 GHz. The SKA-Low array will transfer an average of 9 TB/s of data to the Science Processing Centre (SPC) in Perth while SKA-Mid will transfer approximately 20 TB/s to the SPC in Cape Town, where two supercomputers are located, one at each SPC.
At these centres, the Central Signal Processors (CSPs) will correlate or condition the signals; however, it will not be possible to store all this information. For this reason, the Science Data Processors (SDPs) will process the data, flagging and calibrating them to create images, and delete the visibilities. The size of these images is expected to be up to several TB; SKAO will deliver approximately 700 PB per year of calibrated data to the SKA Regional Centre (SRC) computing facilities. The SRCs will provide the scientific community with access to SKAO data and the necessary processing and storage resources for their scientific exploitation.
International SRC initiatives are collaborating with SKAO to build a network of SKA Regional Centres (SRCNet; \citealp{SKA_DeliveryPlan}), a collaborative and interoperable platform where data will be distributed, allowing users to access and analyse it regardless of their location \citep{bolton_srcnet2023}.

Downloading and processing data on desktop computers will be unfeasible due to their limited memory compared to the size of SKAO data. The SRCNet will provide access to data, computational and storage resources, tools, and scientific support to analyse and exploit SKA data. One of its main challenges lies in conducting data processing, analysis, and visualisation remotely within a distributed computing infrastructure. To approach this challenge, the software stack, defined for a node in the first version of the SRCNet, includes (see \citealp{srcnet0.1_devplan}) common data-related services (implemented by the tool RUCIO\footnote{\url{https://rucio.github.io/documentation/started/concepts/rucio_storage_element/}}), local data parsing services, containerised visualisation tools, an interactive analysis interface (based on JupyterHub), monitoring services, a science platform gateway (based on ESAP; \citealp{ESAP2022}), and a science platform (based on CANFAR; \citealp{CANFAR}).

The SRCNet will embrace the application of Open Science and FAIR principles (Findability, Accessibility, Interoperability, and Reusability; \citealp{FairPrinciples2016,FAIR_Software2022}), which are founding principles of SKAO \citep{SKA_ConstructionProporsal}. These principles are a collection of guidelines on how to publish scientific resources (from raw or processed data to software tools and workflows) relying on computational systems, as it will be required by the high volume of data from next-generation observatories. Those principles have been devised to make science transparent and collaborative, this way reducing inequality and boosting productivity and research visibility, especially for developing countries. The problems of findability and reusability can be tackled by integrating processing, analysis and visualisation services in virtual research environments. These federated environments require being interoperable in order to work with data of different types, also needing applications and workflows. The Virtual Observatory (VO) community is already addressing these challenges \citep{IVOA_ESAP2020} by developing standards that facilitate data and software interoperability.

The VO is a platform or system where astronomical data can be shared, accessed, analysed and visualised. It works as a link between different projects and provides a framework for data centres to supply services. The standards to maintain, improve, and enable the VO are set by the International Virtual Observatory Alliance\footnote{\url{https://ivoa.net/}} (IVOA). Many widely used services in astronomy follow these standards, significantly enhancing their capabilities. In particular, interoperability is achieved directly through the definition and implementation of these standards. Some examples are the Simple Spectral Access (SSA; \citealp{ssa1.1}) and the Table Access Protocol (TAP; \citealp{tap1.1}) for spectral data and table access, respectively, or the Astronomical Data Query Language (ADQL; \citealp{adql2023}), a grammar used to make queries for astronomical data. Other examples will be discussed in Sec. \ref{archive}.

One of the data products of radio observations are spectral datacubes, which provide both spatial (on the celestial plane) and spectral information about the objects of study; hence, these data products are three-dimensional (3D). For spectral line data, the wavelength or frequency can be converted to line-of-sight velocity by estimating the Doppler shift. Historically, spectral cubes were predominantly used in radio astronomy; however, Integral Field Units (IFUs) are producing an increasing amount of 3D data in the optical (e.g., MUSE; \citealp{muse2010} and CALIFA; \citealp{califa2012}), infrared (NIRSpec; \citealp{nirspec2022}), and with forthcoming experiments in X-rays \citep{athena_xifu2018}. To effectively analyse datacubes, 3D visualisation techniques are highly valuable, capitalising on the human ability for pattern recognition \citep{hassanfluke2011Petascale}.

\begin{figure}
    \centering
    \includegraphics[width=0.45\textwidth]{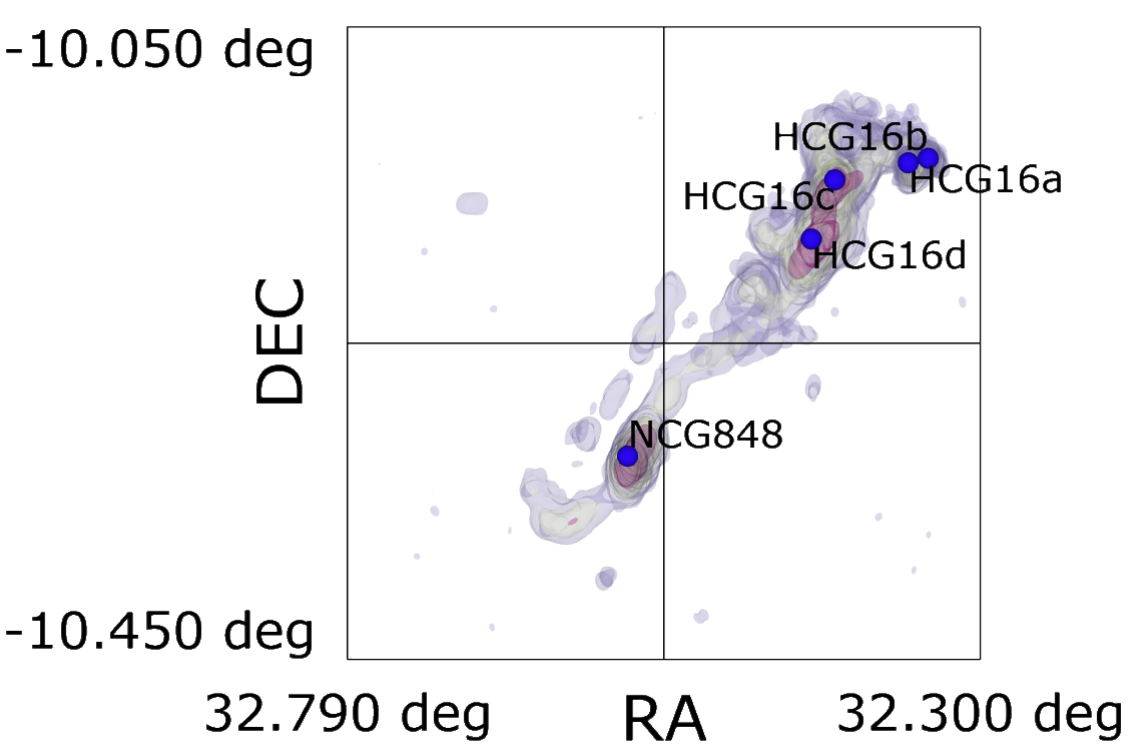}
    \caption{Iso-surface visualisation of HI in HCG16. Blue markers represent group galaxies. The gas shows features caused by the interaction between galaxies, such as the tail between the core of the group and NCG848, and the hook going back. 3D visualisation techniques are used to gain more insight into the dynamics of the gas.}
    \label{fig:hcg16_ra-dec}
\end{figure}

A classical example is the emission line of neutral hydrogen (HI or 21cm line), which can be observed at radio wavelengths and is fundamental for the study of galactic evolution. It is extremely sensitive to galaxy interactions and serves as an excellent kinematics tracer. In particular, we have focused on Hickson Compact Groups (HCGs; \citealp{Hickson1982}). HCGs are small, isolated groups of galaxies characterised by a high surface density and low velocity dispersion. As a result, mergers and strong interactions are common in these groups, making them an ideal environment for the study of galaxy evolution. The HI emission from galaxies in compact groups is often entangled, creating features such as tails and clumps. 3D visualisation techniques are powerful tools for extracting information about the dynamics of the gas; therefore, these HI datacubes are ideal to demonstrate the advantages of 3D rendering software. For example, HCG16, shown in Figure \ref{fig:hcg16_ra-dec}, presents these features, but, even though 2D images can give some information, interactive 3D visualisation techniques are required to perform a complete analysis. In this case, iso-surfaces can be useful, for example, to highlight tails, clumps, or bridges, to separate intra-group gas from disk gas, or to identify disturbed galaxies.

In fact, 3D visualisation techniques are used in many diverse fields. For example, in medical sciences for the analysis of protein structures \citep{protein2020,3Dprotein2019}, in construction for risk management \citep{BIM2019}, in sensor data analytics for advanced driver assistance systems \citep{adas2021}, or to create virtual worlds for gaming or the metaverse \citep{metaverse2023}. In the field of astrophysics, many advances have been made in scientific visualisation \citep{astrovis2021}, for example, in virtual reality \citep{idavie2024}.

However, astronomers predominantly rely on 2D visualisation techniques, such as slicing (either in separate images or video) or moment maps, to extract 3D information. This highlights the need to further develop 3D rendering tools specifically tailored to this field \citep{hassanfluke2011Petascale}. The Hierarchical Progressive Survey (HiPS; \citealp{hips2015}), a tiling scheme for storing and viewing 2D images at multiple resolutions, has been adopted as an IVOA application standard \citep{IVOA_HIPS2017}. This standardisation enhances accessibility and transparency, as demonstrated by services such as Aladin \citep{Aladin2000}. The VO community is actively working on enhancing datacube visualisation with HiPS \citep{hipsgen3d} and Aladin has shown the capability to display slices of a data cube in video mode. However, the lack of established standards for 3D data visualisation is a significant challenge for the implementation of advanced 3D techniques.

In this work, we have conducted a study of existing visualisation software that could be used with astronomical datacubes and made scalable to handle Big Data from observatories such as SKAO. A solution to the challenge of 3D visualisation with Big Data is the creation of 3D models remotely \citep{hassanfluke2011Petascale}, as this approach enables the use of large computing facilities to perform operations that would not be feasible on a desktop computer. Our investigation is focused on the implementation of this solution in a scientific archive following IVOA standards. From this research, we have developed an example archive, deployed within the prototype of the Spanish SRC\footnote{\url{https://ska-spain.es/}} (espSRC; \citealp{espsrc2022}), that transforms spectral line datacubes on the fly into 3D models and renders them through Representational State Transfer (REST) interfaces on the Web. We discuss how IVOA recommendations could be expanded to support this kind of visualisation capability, making the results more accessible, transparent, and reproducible. The purpose of this application is not to replace other visualisation software but to complement it, enhancing our ability to understand and analyse scientific phenomena.

This paper is structured as follows: Sec. \ref{available-tools} describes a selection of available technologies for visualisation whereas Sec. \ref{archive} outlines the 3D visualisation platform, including its integration into the developed archive and a discussion on how IVOA recommendations support visualisation capabilities. Sec. \ref{model-vis} provides a description of the 3D models and the web application created to deliver the visualisation service. The paper concludes with a discussion on the efficiency of the service regarding scalability (Sec. \ref{discussion}) and the conclusions (Sec. \ref{conclusion}).

\section{Visualisation technologies}\label{available-tools}

\begin{table*}[ht]
\centering
\caption{Characteristics of visualisation technologies}
\begin{tabular}{|c|cccccc|}
\hline
\rowcolor[HTML]{C0C0C0} 
\textbf{Technologies}                                            & \textbf{Type} & \textbf{Web support} & \textbf{3D capabilities} & \textbf{\begin{tabular}[c]{@{}c@{}}Analysis\\ capabilities\end{tabular}} & \textbf{Scalability} & \textbf{Field}                                                 \\ \hline
X3D/X3DOM                                                        & Standard      & Yes          & High                & None              & Medium               & General                                                        \\ \hline
\rowcolor[HTML]{EFEFEF} 
\begin{tabular}[c]{@{}c@{}}3D Slicer/\\ SlicerAstro\end{tabular} & App           & Indirectly   & High                & High              & Medium               & \begin{tabular}[c]{@{}c@{}}Medical/\\ Astronomy\end{tabular}   \\ \hline
ParaView                                                         & App/Library   & Indirectly   & High                & Medium            & High                 & \begin{tabular}[c]{@{}c@{}}General/\\ Engineering\end{tabular} \\ \hline
\rowcolor[HTML]{EFEFEF} 
\begin{tabular}[c]{@{}c@{}}Blender/\\ FRELLED\end{tabular}       & App           & Indirectly   & High                & Medium              & Medium               & \begin{tabular}[c]{@{}c@{}}Art/\\ Astronomy\end{tabular}       \\ \hline
Plotly                                                           & Library       & Yes          & High                & None              & Medium               & General                                                        \\ \hline
\rowcolor[HTML]{EFEFEF} 
VisIVO/Vialactea                                                & App           & No           & Medium              & Medium            & High                 & Astronomy                                                      \\ \hline
CARTA                                                            & App           & Yes           & Low                & High            & High                 & Astronomy                                                      \\ \hline
\rowcolor[HTML]{EFEFEF} 
Aladin                                                           & App           & Yes          & Low                & Low            & High                 & Astronomy                                                      \\ \hline
glTF                                                             & Standard      & Yes          & High                & None              & Medium               & General                                                        \\ \hline
\rowcolor[HTML]{EFEFEF} 
\end{tabular}
\label{tab_technologies}
\end{table*}

%\caption*{All technologies are open source and support the visualisation of data in 2D, although some are better suited than others for this purpose. It shows whether they are supported for web browsers (\textit{Indirectly} means that the 3D models can be visualised in a web page, but that it must be exported to a certain format and opened with a particular online viewer), \textit{3D capabilities} describes the ability to create different types of 3D visualisations and their interactivity. \textit{Analysis capabilities} includes options such as computing moment maps and histograms or creating masks. In the \textit{Scalability} column, "High" indicates that the technology is currently capable of handling large datasets, "Medium" signifies that it could manage them with external software and/or superficial modifications, and "Low" denotes that fundamental modifications are necessary to accommodate large datasets.}

This section summarises key features of selected technologies reviewed during our research. The purpose of the review is to identify suitable technologies to develop new visualisation techniques for astronomy, capable of scaling up to meet SKA requirements. Understanding the strengths and limitations of these technologies is crucial for this purpose. The study focuses on open-source or free software available to the community for the visualisation of 3D data. It encompasses tools from various fields, including astrophysics, engineering, medical sciences, and gaming, which also require applications capable of handling large datasets.

After an exploratory search, nine tools or technologies were selected for a deeper review, summarised in Table \ref{tab_technologies}. The evaluation criteria included scalability, the ability to interact with the data (interactivity), 3D rendering capabilities, and integration into a web page. In the \textit{Web Support} column of the table, "Indirectly" indicates that the 3D models can be visualised on a web page, but that they must be exported to a specific format and opened with a particular online viewer. The \textit{3D capabilities} column describes the ability to create different types of 3D visualisations and their interactivity. \textit{Analysis capabilities} includes options such as computing moment maps, generating histograms, or creating masks. In the \textit{Scalability} column, "High" indicates that the technology is currently capable of handling large datasets, "Medium" signifies that it could manage them with external software and/or minor modifications, and "Low" denotes that fundamental modifications are necessary to accommodate large datasets. The following paragraphs describe some noteworthy tools identified during the review.

\subsection{CARTA}

The Cube Analysis and Rendering Tool for Astronomy\footnote{\url{https://cartavis.org/}} (CARTA; \citealp{carta_software}) is a visualisation and analysis tool designed for radio data, focusing on the large file sizes ($\gtrsim$ TB) produced by modern telescopes. It is being developed as a replacement for the CASA viewer \citep{CASA2007} and is attracting many users due to its speed, efficiency, and scalability. CARTA uses a client-server architecture and adopts a new HDF5 schema for datacube visualisation \citep{carta2020}.

Memory resources are usually more limited than storage resources; therefore, CARTA creates HDF5 files that are larger than FITS files but can be read faster and more efficiently. The schema presents a hierarchical structure, storing data at different resolutions as well as other information such as histograms and statistics. This reduces the need to perform several memory-intensive calculations on the fly. In addition, the server does not need loading the entire cube due to a 2D tiling model, making memory requirements almost independent of the cube size. Besides visualisation, it also offers many features for analysis, such as spectral line querying, profile fitting, and a moment map generator.

On the other hand, it does not produce 3D visualisations, although this is a feature that developers plan to add in future releases, according to its roadmap. Furthermore, it does not follow IVOA recommendations, making interoperability more challenging.

\subsection{VisIVO/Vialactea}

VisIVO \citep{visivo2015} encompasses a set of tools and services designed for the VO, focusing on the visualisation and exploration of multidimensional datasets. It includes a stand-alone application, a server platform, and a web portal supporting the server. This client-server architecture is essential for handling very large data volumes. It enables the visualisation of 3D data using volume rendering, iso-surfaces, and slices, implemented via the Visualisation Toolkit\footnote{\url{https://vtk.org/}} (VTK).

The Vialactea Visual Analytics Tool \citep{vialactea2018} is based on VisIVO and combines different types of visualisations to perform a multi-wavelength analysis. It includes galactic plane data managed by the ViaLactea Knowledge Base \citep{vlkb2016}, which can be accessed remotely. Vialactea can interactively visualise 3D data using iso-surfaces and combine it with catalogues or 2D images from different surveys. In addition, it can calculate moment maps, spectral energy distributions, contours, and other data products.

\subsection{3D Slicer/SlicerAstro}

3D Slicer\footnote{\url{https://www.slicer.org/}} is a visualisation and analysis application developed for medical sciences. It supports data up to 4D and can create 3D models, including image segmentation, although the standard application only supports a limited number of medical data types. For other use cases, extensions can be created and accessed through the 3D Slicer App Store. These extensions include various medical use cases, virtual reality, and an extension for astrophysics called SlicerAstro\footnote{\url{https://github.com/Punzo/SlicerAstro/wiki}} \citep{slicerastro2017}, which focuses on the 21 cm line. As it is designed for this specific purpose, it implements many useful features, such as coupled 2D/3D visualisation, interactive filtering, masking, and modelling. 3D Slicer is fully scriptable and can run in a Docker container or in a Jupyter notebook.

However, SlicerAstro is no longer supported and is not included in the latest versions of 3D Slicer. As a result, it may be difficult to install on some devices, and important features, such as cloud computing, might not be available.

\subsection{ParaView}

Another tool outside the field of astronomy, ParaView\footnote{\url{https://www.paraview.org/}} \citep{paraview2005}, is widely used in material sciences, engineering, and medical sciences, among other fields. It has a client-server architecture that makes it highly scalable, although it can also run locally. Its strength lies in its ability to handle large datasets and its flexibility to work with any kind of data, as it includes a scripting interface and uses the VTK engine. In addition, it offers a wide range of analysis capabilities, with more than 200 filters.
However, ParaView is not designed for astrophysics and does not include a built-in reader for FITS files \citep{FITS2010}. As a result, many steps are required to produce a 3D model from astronomical data. Its interface is highly complex and has a steep learning curve; however, this complexity provides greater flexibility in representing data.

\subsection{Plotly}

A more basic library, \textsc{Plotly}\footnote{\url{https://plotly.com/}}, can create interactive graphs in an intuitive way in many programming languages. It supports a wide variety of plots, including interactive 3D plots, and can export them to HTML with features such as animations and buttons. It is widely used; for example, CARTA incorporates it as a third-party library. Iso-surface visualisations of radio datacubes can be generated with just a few lines of code, but the resulting files are extremely large, far exceeding the size of the original data.

\subsection{Blender/FRELLED}

Blender\footnote{\url{https://www.blender.org/}} is a powerful computer graphics software widely used for animations, visual effects, and motion graphics. Although it is primarily designed for artistic purposes and does not natively support astronomy standards, it can still be very useful for processing and visualising data.
The FITS Realtime Explorer of Low Latency in Every Dimension\footnote{\url{http://www.rhysy.net/Code/Software/FRELLED/}} (FRELLED; \citealp{FRELLED2015}) is a FITS viewer implemented in Blender. It creates an interactive 3D view of the data and, since it was designed for spectral lines, provides several tools for source extraction and analysis, including querying compact sources, masking, and generating contours.

\subsection{X3D/The X3D-pathway}

Extensible 3D\footnote{\url{https://www.web3d.org}} ($\mathrm{X3D^{TM}}$) is a royalty-free and open ISO/IEC standard (since 2004) for modelling, viewing, and printing interactive 3D data, developed and maintained by the Web3D Consortium. X3D's longevity ensures long-term stability, reusability, and backwards compatibility. It can store models in various formats, such as XML, JSON, and Compressed Binary Encoding, and is accessible across multiple platforms. X3D is easily integrated into web browsers with X3DOM\footnote{\url{https://www.x3dom.org/}} to enable interactive 3D rendering, with additional interactive features possible through JavaScript (JS).

The X3D-pathway \citep{Vogt2016} is an approach that leverages these frameworks to visualise datacubes across different wavelengths. It has already been utilised by \citet{Vogt2017} with optical data and by \citet{Jones2019hcg16} and \citet{Narumba2021meerkat_debris} with HI data. This approach creates X3D models using the Python library \textsc{MayaVi}\footnote{\url{https://docs.enthought.com/mayavi/mayavi/}} and provides a template HTML file to represent them with interactive options.

\subsection{glTF}

The Graphics Library Transmission Format\footnote{\url{https://www.khronos.org/gltf/}} (glTF) is a royalty-free standard for 3D scenes and models. It was released as an ISO/IEC standard in 2022 and is maintained by the Khronos Group, representing a similar approach to X3D. It stores data primarily in JSON or binary format and has become increasingly popular in recent years due to its efficiency. glTF models are supported by a variety of visualisation tools and can be integrated into web pages using various third-party libraries.

\subsection{Aladin}

Aladin \citep{Aladin2000} is a desktop or web-based tool that includes, among other features, the capability to access and use HiPS services to display multi-wavelength images from various databases at different resolutions. Aladin is compliant with IVOA standards and is interconnected to other visualisation or analysis tools. In addition to displaying 2D images, it also allows users to view 3D datacubes as slices or as a video, and to superimpose information from catalogues or archives such as \textit{Simbad} and \textit{VizieR}.

\subsection{Overview of selected technologies}

Many methods and technologies exist to visualise multidimensional data, ranging from slicing to full volume rendering, and utilising desktop applications, scripting, or web-based tools. This study has highlighted several limitations in the reviewed technologies, which include: 1) lack of 3D rendering capabilities, 2) insufficient interactivity to extract valuable information from the data, 3) limited scalability for Big Data, 4) lack of interoperability, and 5) accessibility challenges (e.g. difficulty with installation or sharing models). For this reason, we have developed a new tool that addresses these drawbacks. Nevertheless, our tool does not incorporate many capabilities found in other reviewed software, making them complementary.

Considering the specific requirements of a scientific archive, we have prioritised tools that can be integrated into web pages. Since there are already strong tools for analysis, we have focused more on producing visualisations with alternative interactive options, such as hiding or showing surfaces and adding markers. In addition, stability and compatibility with various open-source software are advantageous for applications in astrophysics. Among all the technologies examined, we have identified X3D and the X3D-pathway as the most suitable to fulfil these requirements. A detailed explanation is provided in Sec. \ref{model-vis}, which also includes a full description of the method used to generate 3D models with X3D.

\section{3D visualisation platform}\label{archive}
The scientific process with SKAO observations will predominantly be conducted on a distributed and federated platform due to the large scale of data it will produce. This requires scientific archives to evolve from being solely data providers to becoming data and service providers. Current 3D visualisation technologies are unable to handle SKA-scale data, implying that new tools must be integrated into scientific archives to enable effective analysis of astrophysical phenomena.

As explained in Sec. \ref{available-tools}, next-generation spectral cube visualisation software must produce interactive 3D renderings for large datasets, be capable of managing such datasets, and remain accessible to users even with limited computational resources. Reviewed technologies do not fully meet these requirements; therefore, we have developed a new method that addresses these criteria.

Visualisations are generated on a remote server to tackle the scalability issue, and, to improve accessibility, they are delivered to users through a scientific archive. Integration with the web enhances accessibility while providing interactive capabilities. This section describes the deployment of the science archive (Sec. \ref{science-archive}), the integration of a visualisation service within the archive (Sec. \ref{integration}), and the architecture of the system (Sec. \ref{architecture}).

\subsection{Scientific archive}\label{science-archive}

We have deployed the Data Center Helper Suite\footnote{\url{https://docs.g-vo.org/DaCHS/}} (DaCHS; \citealp{dachs2014}) on the espSRC scientific platform to create a scientific archive. DaCHS is an integrated package designed to facilitate the publication of astronomical data and services to the VO and the Web. It supports the entire workflow, from data ingestion to service definition. DaCHS adheres to all major IVOA protocols and recommendations, covering aspects such as data discovery, data access, and registry. It can publish catalogues, spectra, images, or multidimensional datacubes, in addition to services that enable queries, link additional resources, or process data. DaCHS archives are created by writing what are called resource descriptors, which contain information about available services.

Our 3D visualisation service (explained in Sec. \ref{model-vis}) has been implemented in a custom-made DaCHS archive, ensuring compliance with various IVOA standards to make it accessible and interoperable. However, additional development is required to fully integrate a visualisation service within the VO.

\subsection{Visualisation service within the VO}\label{integration}

Key IVOA standards for our research are described below, although most standards are interconnected and, therefore, some cannot be used independently. The IVOA Support Interfaces (VOSI; \citealp{VOSI1.1}) document describes the minimum characteristics a web service needs to be part of the VO, focusing on metadata, while the Data Access Layer Interface (DALI; \citealp{DALI1.1}) defines resources, parameters, and responses common to all Data Access Layer (DAL) web services. VOSI-capabilities are required by every DAL service and consist of an XML document that describes a given service.

The ObsCore Data Model \citep{obscore1.1_2017} includes key parameters to discover data from astronomical observations. It is used by the Simple Image Access protocol (SIA; \citealp{sia2.0}) to provide access to multidimensional images and by ObsTAP \citep{obscore1.1_2017} to enable more flexible access to tables. Datalink \citep{datalink1.1} works with data access protocols to provide access to additional resources connected to the data. The Server-side Operations for Data Access (SODA; \citealp{soda1.0}) defines a REST web service to perform server-side operations and access resulting data.

DAL services can be incorporated into the IVOA architecture in various ways. Our visualisation service is related to DAL services, although it has certain characteristics that make it different (explained in detail in Sec. \ref{standardisation}). Several approaches to incorporate custom DAL services into the VO are discussed in Secs. \ref{vosi}, \ref{sia}, and \ref{datalink}, highlighting how users can access these services. These alternatives have been explored by the authors within the framework of SRCNet development.

\subsubsection{VOSI-capabilities}\label{vosi}

The IVOA Registry \citep{Registry1.1} provides descriptions of data and services (including potential visualisation services) and facilitates the discovery and selection of relevant resources for specific science cases. A visualisation service supporting VOSI-capabilities could be included in the Registry and accessed by providing suitable identifiers.
However, this is not the expected usage strategy because it requires 1) the user to have prior knowledge of proper identifiers, which is not straightforward, and 2) expanding the definitions of elements already present in the standard, which contradicts the recommended approach of simplifying these elements \citep{useofcapabilitiesVO}. Furthermore, this method does not allow the implementation of the visualisation service in a science archive without relying on other IVOA recommendations designed for this purpose, as illustrated in the following examples.

\subsubsection{SIA data discovery}\label{sia}

SIA is used to discover multidimensional data. Query responses are ObsCore tables that contain metadata and references to the dataset. These tables can include a standard Datalink resource that details service parameters, known as a service descriptor, which is provided through a URL.

The service descriptor is an optional resource in the SIA recommendation, but this parameter could be used to provide a visualisation service, noting that it represents a valid solution if it is the only feature to be linked to the dataset. The ObsCore table can provide metadata to determine values for some service parameters, such as the coordinates and bounds of the image, but not for all parameters in some cases. Typically, for each service, there will be a single, general service descriptor that can be used with multiple data identifiers.
This option can be incorporated into an archive; however, the input parameters for the service are not explicitly described in the response, delegating the responsibility on the user to locate them through other mechanisms, such as consulting the service documentation. In addition, it does not allow linking additional resources to the dataset.

\subsubsection{Datalink}\label{datalink}

\begin{figure}[t!]
    \centering
    \includegraphics[width=0.45\textwidth]{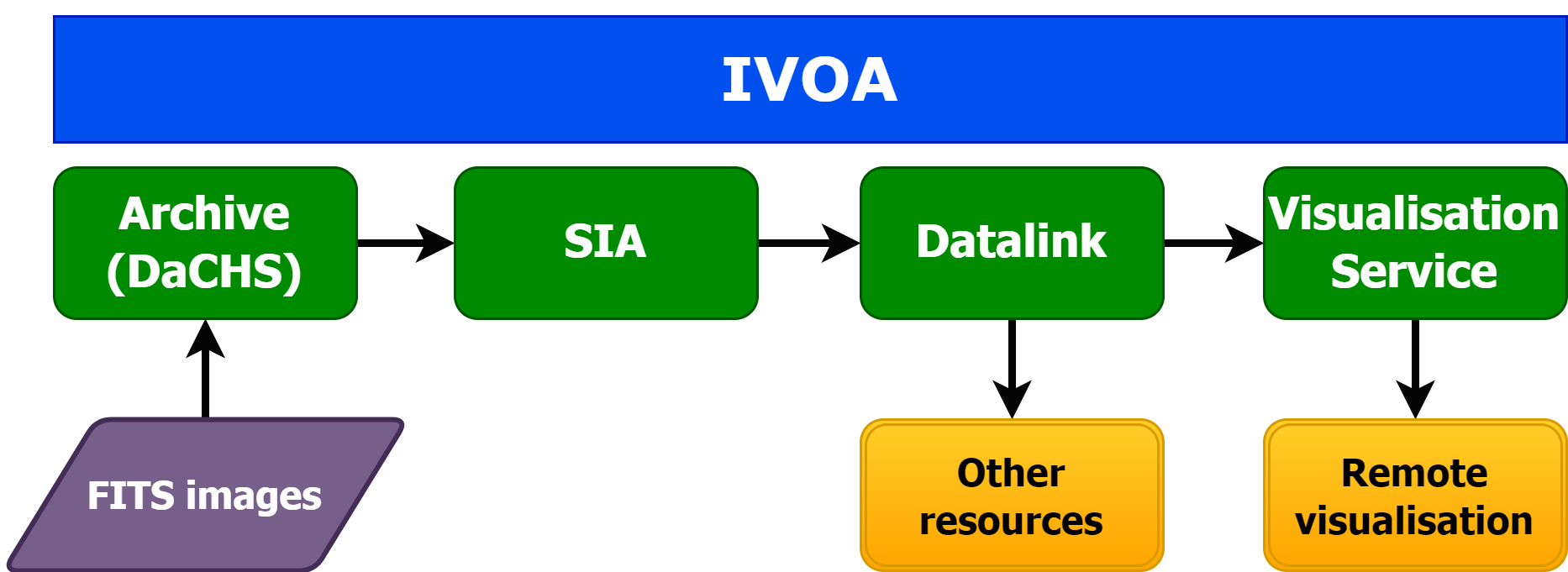}
    \caption{Simplified flow diagram illustrating the process of creating remote visualisations within a scientific archive using DaCHS.}
    \label{fig:service-w/datalink}
\end{figure}

\begin{figure}
    \centering
    \includegraphics[width=0.45\textwidth]{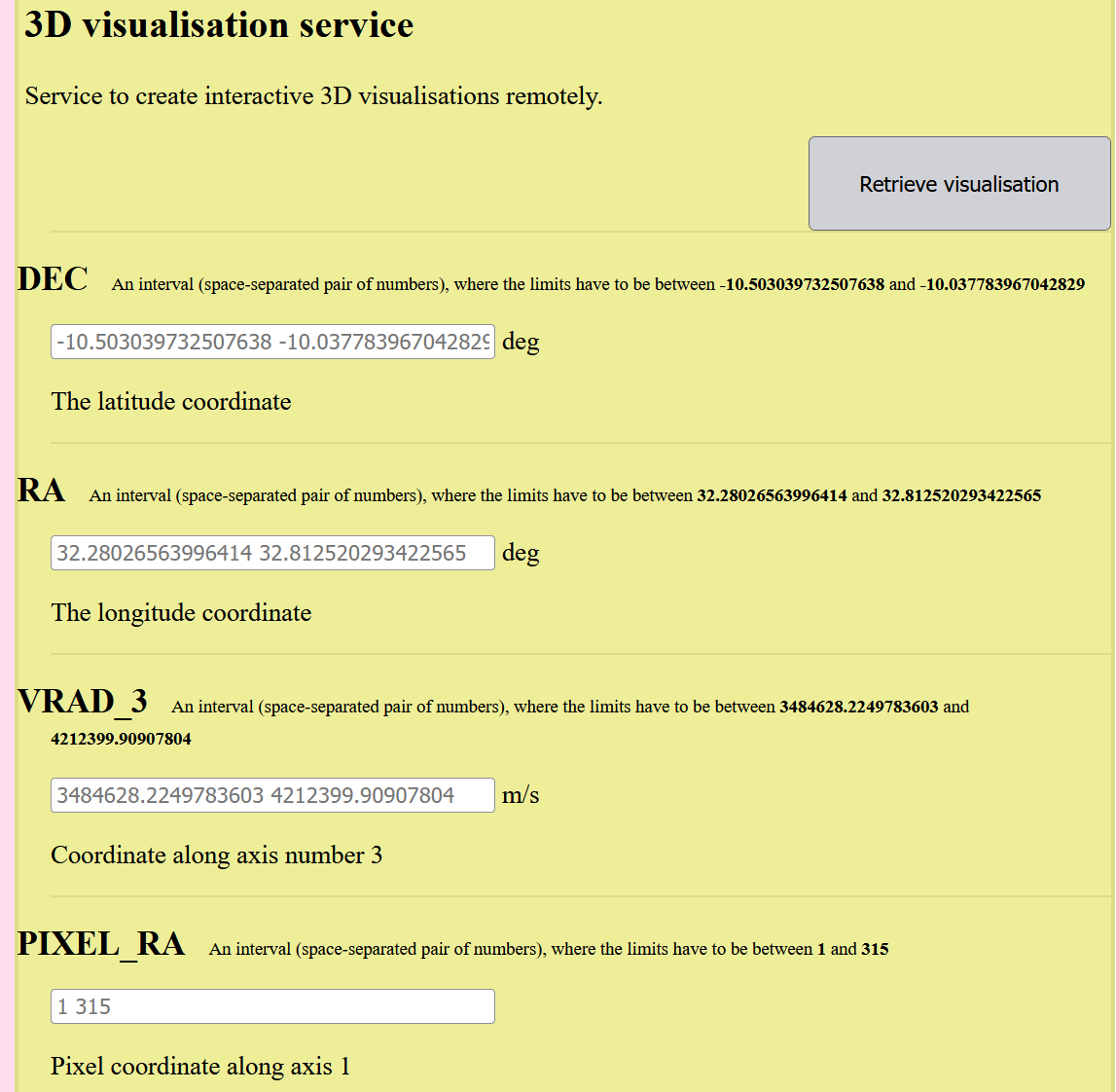}
    \caption{Our deployment of the 3D visualisation service as a REST interface is accessed from a DaCHS archive via Datalink. The service provides options to input parameters, such as coordinates, as well as others not displayed in the figure. X3D models and the custom web application are generated remotely and automatically through this web service.}
    \label{fig:vis_serv_dl}
\end{figure}

This is the alternative that we have chosen, which also involves data discovery responses. It is illustrated in Figure \ref{fig:service-w/datalink} as a flow diagram. In addition to the service descriptor mentioned above, SIA requests can return another feature of Datalink, which is a document containing various static links or dynamic pointers to service descriptors. These service descriptors can include a visualisation service, IVOA services, and/or others.
In this case, valid values should be provided for the required parameters of the service, rather than requiring the user to determine them as in the previous case. This approach allows the visualisation service to be accessed directly from Datalink, which can, in turn, be accessed from an archive. Figure \ref{fig:vis_serv_dl} shows an example of the REST interface including the visualisation service, where the user can input parameters to customise visualisations.

\subsubsection{Standardisation of visualisation services}\label{standardisation}

Having standards for services focused on server-side processing is essential for future infrastructures like the SRCNet, as they enhance the implementation, accessibility, and interoperability of services. SODA is the IVOA standard for server-side processing and must include certain resources: at least one of \{sync\} or \{async\} resources conforming to the DALI description, VOSI-capabilities, and VOSI-availability.
In addition, SODA prescribes a number of standard parameters (e.g., id, pos, circle, etc.), which, along with other custom parameters that can be implemented, must adhere to DALI specifications. The main use case of SODA is to retrieve cutouts from images or spectra; however, \citet{soda1.0} explain that SODA has been devised to provide access to data and server-side processing. Therefore, it could be inferred that a service supporting this standard could also be used to create visualisations.
On the other hand, the fact that SODA is located within the data access layer of the IVOA architecture suggests that its authors intended it to provide access only to data. However, apart from this, the technical document does not impose any explicit limitation on the SODA response. This implies that, theoretically, a SODA service could provide access not only to FITS files but also to operations or applications.

An approach for our service to conform to SODA would be to decouple the processing operation from the application. The service would return only the X3D file and the visualisation would be provided by a separate software. As a result, the service would solely transfer data (in alignment with the intent of the IVOA recommendation), and the visualisation, along with its interactive features, would be delivered by a third-party application. At the time of writing, this option is being tested by the espSRC team in collaboration with SKAO and SRCNet teams.

Future computing infrastructures will need to provide server-side operations and applications capable of processing, analysing, and visualising data due to the large size of the datasets. Standardisation is fundamental to maintain interoperability between services provided by new facilities. We have identified two potential approaches to standardise services that grant access to applications or operations. First, SODA could be adapted to explicitly state its validity for this purpose, as it already includes several common characteristics. In the context of science platforms, operations and applications can be considered a type of data. By adopting this broader interpretation of data, an application could theoretically be served using the current version of SODA (1.0). This adaptation could involve, for example, adding a use case to the existing recommendation. Second, a new standard could be developed to implement these services, such as Server-side Operations for Processing Access (SOPA). This approach is also reasonable, as operations and applications are inherently more complex than traditional data formats such as images or spectra.

Although service descriptors and IVOA standards such as UWS are already contributing to partially addressing this problem, SOPA could define a high-level specification for the discovery and access of operations not necessarily exposed as web services. This standardisation would focus on minimising the transfer of large data volumes and facilitating not only access to applications but also the movement of software to the data.
For this reason, we envisage that some of the mandatory parameters SOPA would require include: the type of operation to perform on the data (e.g., visualisation, source extraction, application of machine learning techniques, changing resolution, removing noise, etc.), the input data type (e.g., datacubes, images, time series, or any existing IVOA data type), and the output data type (the data types resulting from this operation).
As an example, other optional parameters could specify which science platform exposes the operation, whether the operation can only be executed on that platform, or if it can be transferred to another science platform.

We envisage that SOPA could also define some key parameters for its output, e.g., a reference to the science platform hosting the operation and its operation ID, input parameters required to invoke the operation (making use of other IVOA standards such as UCD descriptors), or information on how to access the operation (e.g., through containers, virtual machines (VM), service descriptors, or any other technology enabling the exposure of operations on data).

We have developed a proof of concept by modifying the standard implementation of SODA in DaCHS to enable access to applications. Nevertheless, both approaches appear valid to standardise services that provide either operations or access to applications. It is important to emphasise that any standardisation—whether by expanding the scope of SODA or by creating a new standard—would require consensus from the IVOA community.

\subsection{System architecture}\label{architecture}

\begin{figure*}
    \centering
    \includegraphics[width=0.95\textwidth]{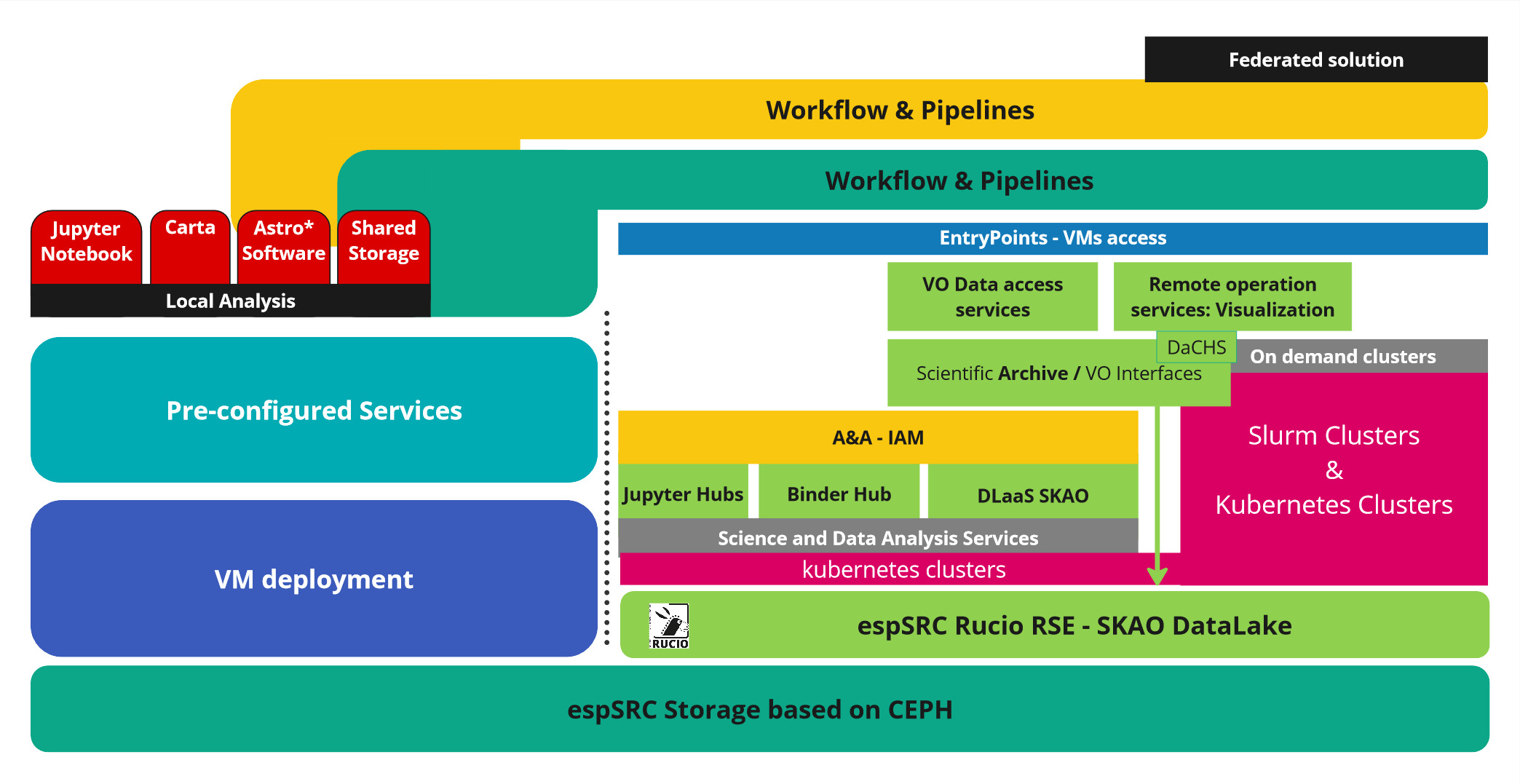}
    \caption{Architecture of the espSRC as a federated infrastructure. Cloud services such as VM, \textit{Slurm} clusters or data analysis services are deployed on top of the cloud storage system. Upper layers include authentication services or a science archive.}
    \label{fig:architecture}
\end{figure*}

The prototype of the Spanish node of the SRCNet, called espSRC, is being developed at the Instituto de Astrofísica de Andalucía (IAA-CSIC). It includes a cloud computing infrastructure that provides support for scientist working with SKA precursors and pathfinders, for SKAO Data Challenges and for training activities such as the IAA-CSIC Severo Ochoa SKA Open Science School (2023).

The espSRC hosts, at the time of writing, an OpenStack cloud computing platform equipped with 240 physical cores, 2.9 TB of RAM, 1.7 PB of raw storage managed by Ceph, and provides computing and storage resources and services for various scientific projects. In terms of connectivity, the computation and storage components are interconnected via a 100 Gbps network, and the cluster features a 10 Gbps connection to the internet. Regarding storage management, the read/write ratio is optimised using solid-state drives (SSDs), enabling operations directly on the server before data transfer.

DaCHS has been deployed on a VM instantiated within the espSRC platform, equipped with 4 CPU cores and 8 GB of RAM. Additionally, for storage, the VM is configured with a distributed block storage unit providing a total of 500 GB dedicated to data management operations with the DaCHS service. In terms of connectivity and networking, the VM is accessible through internet via an entry point connected to the service provided by DaCHS.

The system architecture, as depicted in Figure \ref{fig:architecture}, comprises multiple tiers of resources, including computing, storage, and networking. At the base of the diagram is the espSRC storage system, upon which all cloud services are mounted. These services range from the provisioning of VMs to services for the distribution and management of data within a RUCIO datalake, as well as individual services for VMs and more complex services for interactive data analysis and batch processing within a \textit{Slurm}\footnote{\url{https://slurm.schedmd.com/overview.html}} cluster. The upper layers incorporate authentication services, alongside access to VM services and other data analysis services, including specific components for VO data access, visualisation, and scientific archive access. These developments are partly conducted by the teams participating in the SRCNet prototyping and development phases.

In its current configuration, operations like data transformation into 3D models are conducted within the VM. However, transitioning to more flexible computation environments like platforms based on containers (e.g., \textit{Kubernetes}\footnote{\url{https://kubernetes.io/docs/concepts/overview/}} clusters) or other platforms that use scheduling systems and workload managers (e.g., \textit{Slurm}), could lead to significant improvements in scalability, availability, and fault tolerance. 
This shift enables the dynamic allocation of hardware resources based on service demand, ensuring optimal performance even during peak usage periods. Indeed, the espSRC can support \textit{Slurm} for certain distributed computing processes and dynamically adjust resources based on specific process needs. Moreover, the ongoing migration of visualisation software to containers underscores the technical evolution underway, although this particular upgrade is out of the scope of this article.

The architecture has been adapted to enable data access (e.g. images, catalogues, spectra) via DAL protocols, as well as to support more advanced operations that allow remote processing without requiring data downloads, as exemplified by the visualisation service. The procedure implemented by the service involves transforming spectral datacubes into 3D models and creating an application to visualise them, as detailed in Sec. \ref{model-vis}.

\section{3D modelling and visualisation}\label{model-vis}

Proper examination and analysis of spectral datacubes require 3D visualisation techniques in many scientific cases. However, 3D tools are less developed than their 2D counterparts due to 3D data being historically less common and to the additional complexity of rendering 3D models, highlighting the need for further research in this field. Volume rendering is the most straightforward technique to represent 3D data and is used by many tools. This method assigns colour and opacity values to every voxel and creates a 2D projection of the 3D dataset, based on a viewpoint relative to the volume. Another approach is surface rendering, which involves extracting iso-surfaces (surfaces with uniform intensity) and representing them using polygon meshes. Polygon meshes consist of vertices and edges that form polygonal faces, typically triangles, which collectively define the surface of 3D objects. In general, volume rendering provides more detail about the global structure and gradual transitions, while iso-surface rendering highlights relevant features such as correlations and boundary regions.

We have chosen to represent data using iso-surfaces in our visualisation software because we are focusing on the visualisation of HI data and extended large structures, as explained below. Based on this approach, the file size of iso-surface models is generally smaller, as it stores only the surface points rather than the entire data volume. This reduction in file size helps address the challenges associated with Big Data.
In our case, the computation of the iso-surfaces is done as a server operation.
In addition, multiple iso-surfaces can be represented simultaneously, allowing the inner structure of the data to be visualised without visual clutter by simply hiding certain surfaces. Exploring iso-surfaces is often more accessible to non-experts in visualisation because the interaction is more intuitive; for instance, many volume rendering tools require the definition of transfer functions, which can be both confusing and time-consuming. This type of visualisation is also highly valuable for experts studying extended structures, offering better support for the scientific cases we have selected. It is, therefore, a complementary visualisation approach to those provided by other tools.

Our visualisation service is designed for any spectral line datacube, although we have focused particularly on HI observations. Most of the data used for testing involve HCGs, as these objects are expected to exhibit extended structures (see Sec. \ref{introduction}).

The service has been implemented in a science archive to enable remote operations and consists of two layers. The processing layer transforms datacubes into 3D models, while the application layer generates web applications to visualise the models. The exploration described in Sec. \ref{available-tools} led us to select X3D as the standard for storing the models, along with HTML, X3DOM, and JS for creating interactive visualisations on the web.
These technologies were chosen because they: 1) allow sharing and viewing 3D models online without requiring additional software installations; 2) are open-source, long-lasting ISO/IEC standards, ensuring long-term stability and backwards compatibility; 3) provide great flexibility for implementing interactive features, and 4) have the potential to be scalable for Big Data. We selected the text-based XML format of X3D for the models, enabling autonomous model editing and direct implementation in HTML. It should be noted that the X3D model and the web application are generated dynamically; both processes can run independently and in any order, provided that the X3D objects are properly referenced by the web interface. This is useful to connect different application layers (in addition to our web application) to a single processing layer, enabling the 3D models to be used in multiple ways simultaneously. For instance, X3D models can be imported into Blender to create animations.

Both layers are encapsulated in a Python library called \textsc{ViSL3D}\footnote{\url{https://github.com/ixakalabadie/ViSL3D}} (Visualisation of Spectral Lines in 3D; \citealp{visl3d}), which is imported into DaCHS to operate on archive data. The visualisation service is defined within the DaCHS resource descriptor via a service descriptor delivered by the DataLink service, which is linked to an SIA service. The visualisation service has two primary elements that characterise its functionalities: \textit{metaMaker}s, which define the input parameters users can enter, and \textit{dataFunction}s, which define the operations to be performed. DaCHS includes several of these elements prepared for SODA services. We have used some of them for the visualisation service, as they describe the same parameters.

The parameters defined for the service include the coordinate ranges of the cube along each axis, the range of pixels along each axis, the unit used to represent the cube, the values of the iso-surfaces, the survey from which to obtain a 2D image, and the resolution of the 3D model, among others. Each parameter is optional and defaults to a predefined value if not specified by the user.

\begin{figure}[ht]
    \centering
    \includegraphics[width=0.35\textwidth]{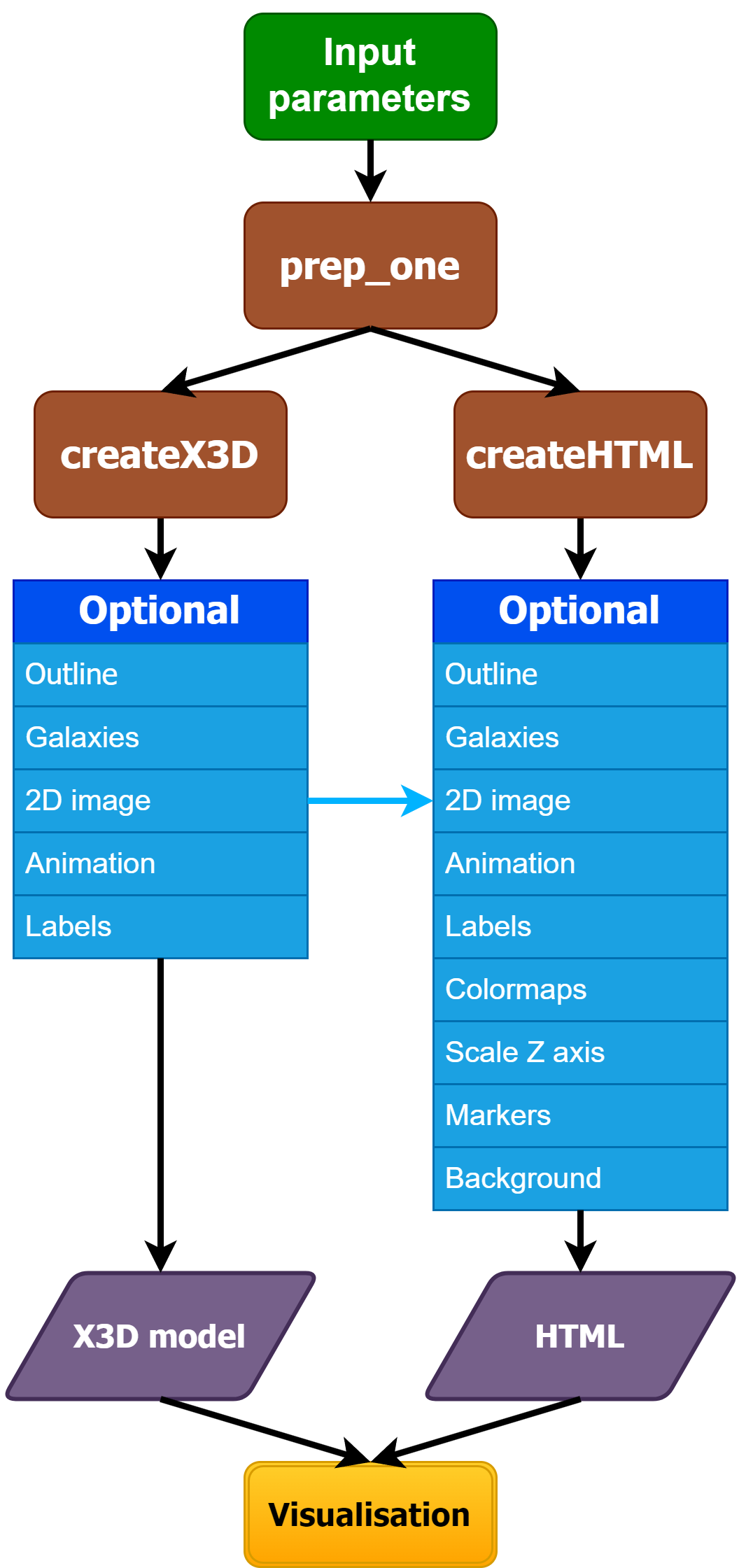}
    \caption{Flow diagram of the visualisation generation process. Input parameters default to the cube’s predefined values if not specified by the user. \textit{prep\_one}, \textit{createX3D}, and \textit{createHTML} are functions of \textsc{ViSL3D}; the first is used to pre-process the data, while the latter two create the HTML and X3D files, respectively.  Some optional elements are connected and cannot be used in the HTML without their equivalent in the X3D model.}
    \label{fig:library}
\end{figure}

\subsection{Processing layer - Model generation}
Our approach to generating and visualising datacubes builds upon the X3D-pathway, described in Sec. \ref{available-tools}. Using this method, 3D models are created in X3D, a format that is highly interoperable and versatile, allowing it to be utilised with various technologies across multiple fields. It can comprehensively represent diverse types of 3D data and it supports a wide range of features, including different shapes, lighting, transformations, materials, navigation, animations, and more.
We have enhanced the X3D-pathway to improve scalability, integrate it into a science archive, and provide more advanced capabilities tailored to the demands of modern next- generation telescopes.

The X3D-pathway employs the Python library \textsc{MayaVi} to create and export 3D models to an X3D file. In this approach, X3D models represent iso-surfaces as a triangle mesh. The primary drawback of using \textsc{MayaVi} is that the transformation from FITS to X3D, performed with VTK, is inefficient, resulting in large file sizes.
The triangle mesh consists of numerous triangular faces, which \textsc{MayaVi} defines using indices, vertex coordinates, normal vectors, and colours, all of which share the same shape. In our case, since the colour remains constant for each triangle in a surface, this information can be omitted without any loss of detail. By excluding this redundant data, the file size is reduced by an average of 25\%. This percentage depends on the geometry of the data and on additional information of the model (labels, number of iso-surfaces...), although the variation is negligible.

The same applies to normal vectors; their calculation for simple models is not computationally expensive to perform on the fly and they can also be removed to obtain files \%25 lighter. Moreover, all numerical values included in the file (including many integers) are formatted in scientific notation with 17 decimal places. Given that the number of numerical values in the files normally exceeds $10^6$, changing the precision and how the values are represented can save a substantial amount of storage space, up to 50\% if the model includes normals and colours. By applying all these optimisations, the efficiency of the transformation process is significantly improved, resulting in X3D files approximately \%80 smaller than the originals.

Another issue is that models created with \textsc{MayaVi} require significant modifications to be rendered properly on a web page and to include interactive options. For instance, the lighting is often inadequate, and objects lack IDs, which must be added or corrected manually outside the \textsc{MayaVi} environment. Additionally, the HTML file required to display the X3D model must be created independently and customised for each model. These limitations, combined with the requirement to generate visualisations for a science archive in an automated manner, prompted us to develop a Python library capable of producing X3D models and their corresponding HTML files, bypassing the need of \textsc{MayaVi} or other tools. Furthermore, we have incorporated numerous features not available in the X3D-pathway, enhancing both the interactivity and utility of the visualisations (see Sec. \ref{application}).

The workflow to create visualisations through the DaCHS service with \textsc{ViSL3D} is presented in Figure \ref{fig:library}. It shows that the library comprises two primary functions: \textit{createX3D}, which generates the X3D model, and \textit{createHTML}, which creates the corresponding web application (see Sec. \ref{application}). These functions require data and coordinates in a specific format, as well as parameters provided through the service. This can be done with the following functions: \textit{prep\_one} (as in Fig. \ref{fig:library}), which is used to create visualisations of a single spectral line and is the one implemented in DaCHS; \textit{prep\_mult} designed for multiple spectral lines; and \textit{prep\_overlay} to overlay multiple spectral lines. There are optional features available for both the X3D model and the web application, some of them connected to each other.

3D models are created using the class \textit{createX3D}, which includes methods to construct various X3D elements. The most critical elements of a model are the iso-surfaces, which are generated using the marching cubes algorithm \citep{marching2003}, implemented in \textsc{scikit-image}. This algorithm generates a triangular mesh from 3D volumetric data for specified iso-values, returning the coordinates of vertices, reference indices for these vertices, the normal vector for each face, and the maximum data value near each vertex.
The vertices and their reference indices are written into the X3D element \textit{IndexedFaceSet} to create the iso-surfaces. Each surface is contained within a separate X3D element. If the anticipated size of the model exceeds a predefined limit, the cube is divided into two or more subcubes, and iso-surfaces are calculated for each; this process is further detailed in Sec. \ref{discussion}.
Additionally, the library includes an option to combine two different datacubes and superimpose them, which is particularly useful for visualising different spectral lines.

Iso-surfaces are the only required element in the model; however, we have enhanced its capabilities and flexibility by incorporating additional elements. For instance, an outline, grids, and labels, can be added to facilitate the localisation of data within the coordinate space.
Additionally, there is an option to include markers for galaxies retrieved from catalogue queries using \textsc{Astroquery}. This can be achieved either by providing the names of galaxies of interest or by performing a general query within the cube’s coordinates. A 2D image, also accessed via \textsc{Astroquery}, can be included in the background to enable a multi-wavelength analysis of the data. The survey from which the image is obtained is specified as an input parameter in the visualisation service.
It should be noted that this last feature could significantly increase the size of the X3D model, depending on the image’s field of view.

Custom parameters of the visualisation service are essential for ensuring scalability to Big Data. Specifying coordinates to create cutouts of the image prior to model generation and defining a resolution for the calculation of iso-surfaces using the marching cubes algorithm can significantly impact the size of the resulting model. This topic is further discussed in Sec. \ref{discussion}.

\subsection{Application layer}\label{application}

\begin{table*}[ht]
\centering
\caption{Objects and interactive options of ViSL3D}
\begin{tabular}{|l|l|}
\hline
\rowcolor[HTML]{C0C0C0} 
Information                                           & Interactive options                                                                                     \\ \hline
                                                      & Rotate, pan, zoom and change centre of rotation of the whole model                                      \\ \cline{2-2} 
                                                      & Show an animation of the 3D model rotating along one axis                                               \\ \cline{2-2} 
                                                      & * Multiple spectral lines in the same visualisation (only for models created in Python)                                                   \\ \cline{2-2}
                                                      & * Make a cutout of the original cube                                                  \\ \cline{2-2}
\multirow{-5}{*}{General}                             & Change background color                                                                                 \\ \hline
\rowcolor[HTML]{EFEFEF}                               & Hide/show iso-surfaces                                                                            \\ \cline{2-2} 
\rowcolor[HTML]{EFEFEF}                               & Change the colormap, including limits and scale                                                         \\ \cline{2-2} 
\rowcolor[HTML]{EFEFEF}
\multirow{-3}{*}{\cellcolor[HTML]{EFEFEF}Iso-surfaces} & * Set the resolution of iso-surfaces                                            \\ \hline 
Viewpoints                                            & Switch among 3 orthografic viewpoints (RA-DEC, Z-DEC, Z-RA)                                             \\ \hline
\rowcolor[HTML]{EFEFEF}                               & Choose coordinates as either the difference from the centre of the cube, with respect to earth, or none \\ \cline{2-2} 
\rowcolor[HTML]{EFEFEF}
\multirow{-2}{*}{\cellcolor[HTML]{EFEFEF}Axes}                                & Change the scale of the Z axis                                                                          \\ \hline 
                                                      & * Display 2D plane, either with a solid colour or an image from the VO       \\ \cline{2-2} 
                                                      & Hide/show plane                                       \\ \cline{2-2} 
\multirow{-3}{*}{2D plane}    & Move the image along the Z axis and display its position                                            \\ \hline
\rowcolor[HTML]{EFEFEF}                               & Add spheres and labels at positions of galaxies from a catalogue                                        \\ \cline{2-2} 
\rowcolor[HTML]{EFEFEF}                               & Hide/show spheres and labels                                                                            \\ \cline{2-2} 
\rowcolor[HTML]{EFEFEF}
\multirow{-3}{*}{\cellcolor[HTML]{EFEFEF}Galaxies}    & Change sphere and label size                                                                            \\ \hline
Markers                                               & Add markers (spheres, tubes, boxes, cones) and their labels                                                                    \\ \hline
\end{tabular}
\caption*{* indicates that this option is only available at the time of creating the X3D model, not in the web application.}
\label{tab_interactivity}
\end{table*}

X3D models are highly interoperable, meaning they can be opened with a variety of software, including desktop and web applications. However, web browsers offer greater potential for achieving both high accessibility and high interactivity. Our X3D models are integrated into HTML files to create interactive 3D visualisations. This integration is particularly straightforward when using X3DOM, an open-source framework for embedding 3D scenes into HTML via a JS library and a CSS stylesheet.
X3DOM utilises the Document Object Model (DOM) interface to make 3D visualisations interactive while supporting standard HTML elements on 3D objects. By default, X3DOM enables users to rotate, zoom, and pan the 3D model, as well as set the centre of rotation. Additionally, it allows for the creation of custom functionalities that directly transform the X3D model through the web application.

When using X3DOM, models stored as external files must be accessed through a server, such as Apache\footnote{\url{https://httpd.apache.org/}}. This requirement poses no issue for a web service in a scientific archive, as servers are the standard operational method in such environments. To enhance usability, we have implemented an option to embed the X3D model directly into the HTML using \textit{createVis}, enabling local use without the need for a server. However, this approach is less interoperable than maintaining the model in a separate file.

The web application is built with the class \textit{createHTML} (Figure \ref{fig:library}) and includes links to X3DOM and other scripts (\textit{LaTeXMathML}, \textit{jQuery} and \textit{js-colormaps}). In this case, all JS functions are optional, having the possibility to display just the X3D model with X3DOM. However, the true strength of the application lies in its enhanced interactive features. A summary of these functionalities is presented in Table \ref{tab_interactivity}.

\begin{figure*}[ht]
    \centering
    \includegraphics[width=\textwidth]{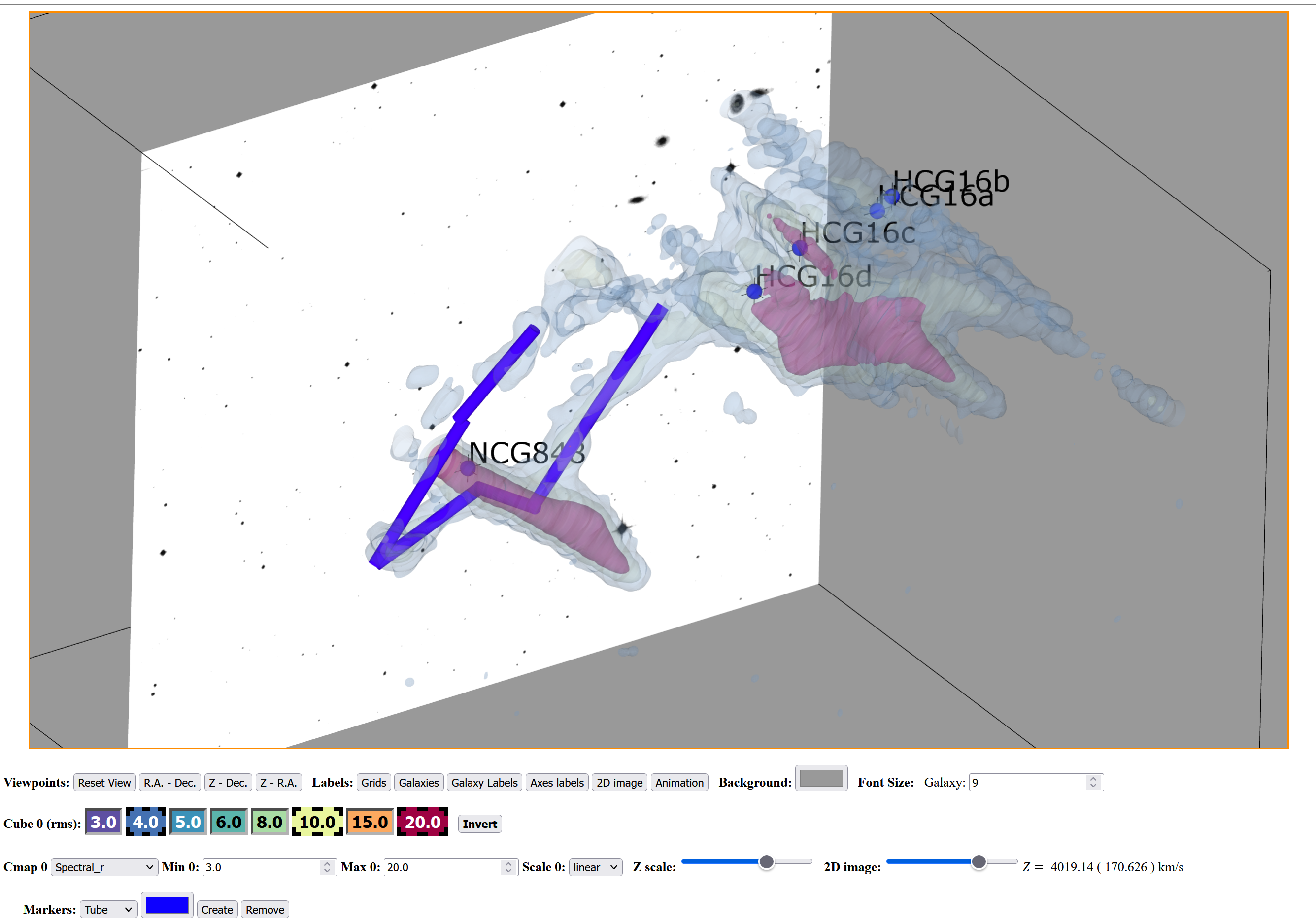}
    \caption{Web application rendering of a 3D model of HI gas in HCG16 observed by MeerKAT. Three out of eight iso-surfaces are displayed along an optical image from DSS2 Blue. Markers for galaxies are shown at positions obtained from a catalogue as well as a tube highlighting a tidal tail. This visualisation has been made online accessing the 3D visualisation service through the DaCHS archive.}
    \label{fig:HCG16}
\end{figure*}

An example of the web interface is shown in Figure \ref{fig:HCG16}, which highlights tidal tails, bridges and clumps of disturbed gas. These interactive visualisations can be included in web pages and accessed via web browsers\footnote{\url{https://amiga.iaa.csic.es/x3d-menu/}}. This particular model represents the atomic gas in HCG 16 as observed by MeerKAT. It contains eight iso-surfaces, of which five are hidden to provide a clearer view. A blue tube has been added to indicate a tidal tail, and a 2D optical image from DSS2 Blue covering the same field of view is superimposed. Additionally, the positions of several known galaxies are marked.

\section{Discussion}\label{discussion}

One of the primary challenges posed by the SKAO is managing the vast data volume it will generate, requiring scalable analysis techniques and visualisation software. The strategy of our visualisation service to address this challenge involves performing data operations remotely on a server and transferring the processed 3D models to the user. We are using the espSRC scientific platform during the SRCNet developing phase as a laboratory for this research as it will facilitate the exploration of how to extend this service to a federated platform such as the SRCNet, once it is operational.

Remote processing enables the use of computational resources from a cloud infrastructure, supporting operations on Big Data. However, users are still required to load and render the 3D models locally, a task that can be highly demanding for heavy models, particularly in terms of Random-Access Memory (RAM). The following subsections examine the scalability of the service, the size of 3D models, their impact on RAM, and the level of interactivity provided.

We have tested the software with a sample of 37 datacubes from the VLA \citep{VLAHCG} and six from MeerKAT (presented in an accompanying paper) to study the relationship between datacubes, the resulting X3D models, and their memory consumption during visualisation. All datacubes are observations of HI emission in HCGs. 

From this data, we generated X3D models with varying resolutions, adhering to common requirements for scientific analysis, e.g., between one and ten iso-surfaces, most of them higher but close to $S/N = 3$. Other types of data, such as IFU data, have also been tested with successful results; however, this discussion focuses exclusively on radio data.

\subsection{Model size}
\begin{figure}     
    \centering
    \includegraphics[width=0.45\textwidth]{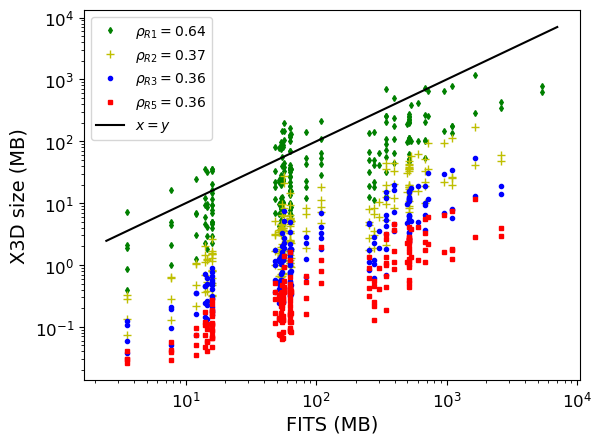}
    \caption{Size of X3D models with respect to the size of FITS files for different resolutions $Rn$ with $n$ the step size. The Pearson's correlation coefficient $\rho$ is shown for each resolution, being $\rho_{R1}=0.64$ for full resolution models and $\approx 0.36$ for the rest.}
    \label{fig:fits-x3d}
\end{figure}

Figure \ref{fig:fits-x3d} compares the sizes of X3D models at various resolutions to their corresponding FITS datacubes. The sizes of the FITS files are grouped into three categories, as data acquired using the same observing mode tends to be of similar size, despite differing content. Consequently, the Pearson’s correlation coefficient ($\rho$) reveals only a moderate correlation ($\rho<0.65$).

\begin{figure}     
    \centering
    \includegraphics[width=0.45\textwidth]{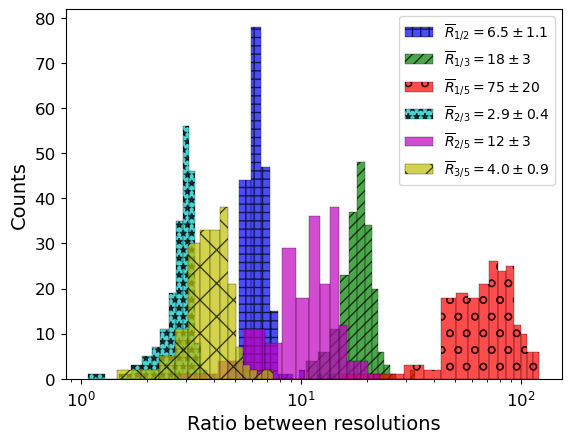}
    \caption{Ratio between the sizes of X3D models of different resolution, e.g., $\bar{R}_{1/3}$ is the ratio between the size of models with full resolution and models with step size equal to 3. The mean and standard deviation of each ratio have been determined.}
    \label{fig:histo}
\end{figure}

Most X3D models are smaller than their corresponding FITS files, even at full resolution. Exceptions occur when iso-surfaces are generated with very low signal-to-noise ratios ($S/N < 3$). Low-resolution models are significantly smaller than full-resolution ones, with differences reaching up to two orders of magnitude in some cases. The resolution is determined by the step size of the marching cubes algorithm; for example, a step size of 1 corresponds to full resolution, while higher step sizes produce lower-resolution models. This relationship is illustrated in Figure \ref{fig:histo}, which shows the ratios of model sizes at different resolutions. Increasing the step size from 1 to 2 results in files 6.5 times lighter on average while changing it from 1 to 5 reduces the size of the models 75 times. The plot also shows that there is more uncertainty when the difference in resolution is larger. 
While using low resolution results in a significant loss of information, it is highly effective for generating preliminary visualisations of large datacubes, enabling the identification of interesting regions for further examination at higher resolutions.

\begin{figure}[ht]
    \centering
    \includegraphics[width=0.45\textwidth]{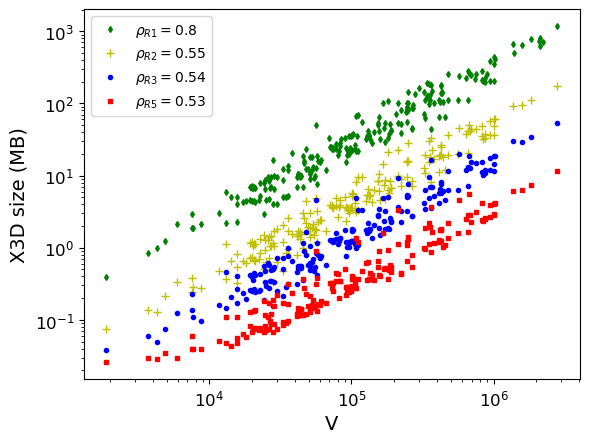}
    \caption{Size of X3D models with respect to the number of voxels higher than the value of the lowest iso-surface $V$, for different resolutions $Rn$ with $n$ the step size. The Pearson's correlation coefficient $\rho$ is shown for each resolution, being $\rho_{R1}=0.8$ for full resolution models and $\approx 0.54$ for the rest.}
    \label{fig:voxels-x3d}
\end{figure}

Although the sizes of FITS datacubes and X3D models are correlated, as shown in Figure \ref{fig:fits-x3d}, the correlation is stronger when comparing model sizes to cube contents. This is because our software generates iso-surfaces rather than visualising the entire datacube. The number of voxels above the lowest iso-surface value is represented against the size of the models in Figure \ref{fig:voxels-x3d}. It should be noted that this plot includes models with different number of surfaces, therefore, some scatter is expected. Nonetheless, $\rho$ is higher in this case compared to Figure \ref{fig:fits-x3d} for each resolution, although full resolution data still exhibits the strongest  correlation.
The reason could be that at lower resolutions the marching cubes algorithm skips a different amount of data in each cube because of the randomness of noise. The trend in the figure is not exactly linear because there is a minimum size X3D models can have. Specially at low resolutions, Figure \ref{fig:voxels-x3d} shows that the functions turn flat towards a low number of voxels. 
The percentage of voxels with $S/N = 3$ is $0.2\pm0.1 \%$ with a maximum value of $0.6\%$. Some cubes in the sample exhibit higher percentages but were excluded due to the presence of artefacts. These findings indicate that most of the volume in the datacubes is either empty or contains only noise; suggesting that selecting specific regions of interest within a larger cube can significantly reduce the size of visualisations without any loss of meaningful information.

\subsection{RAM usage}

\begin{figure}
    \centering
    \includegraphics[width=0.45\textwidth]{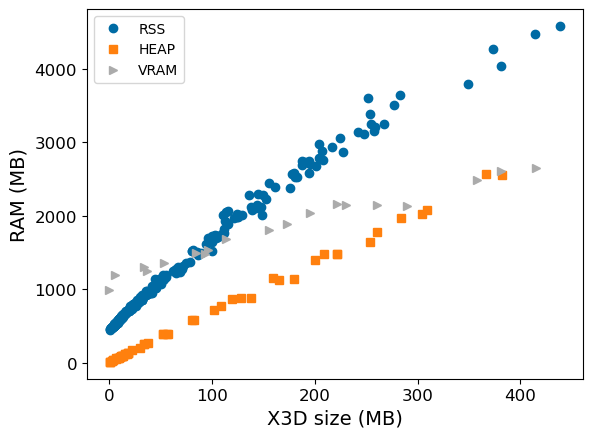}
    \caption{RSS RAM, HEAP memory and GPU VRAM usage of X3D models rendered in the web against the size of those models. All measurements increase with the size of the models as expected. The increase in RSS becomes slower for larger models. Heavier models ( $\gtrsim 400$ MB) have been omitted because they consume $\approx 5000$MB of RSS RAM and result in an \textit{Out of Memory} error.}
     \label{fig:x3d-ram}
\end{figure}

Remote operations can alleviate the problem of transforming very large datasets into 3D models. Even so, with our visualisation service, rendering is performed by the client. This could be a limitation since rendering large 3D objects is highly memory intensive, and standard computers have a limited processing capacity compared to dedicated computing facilities. In our web application, X3D models are loaded and rendered using X3DOM, which uses Graphics Processing Unit (GPU) resources when available. We have measured the memory usage of various X3D models during rendering in the application. Figure \ref{fig:x3d-ram} illustrates the usage of resident set size (RSS) RAM, heap memory allocation (HEAP), and GPU video VRAM (VRAM) as a function of the size of the X3D models\footnote{Rendering has been done in Google Chrome 122.0.6261.129, using X3DOM 1.8.0 and a Dell Alienware 15 R2 with 32GB of RAM (DDR4) and an Intel i9 processor.}. RSS represents the total memory of the process except VRAM and swap, but including child processes created by the browser, HEAP, libraries, rendering engine overhead, cache, etc. HEAP is the memory of JS objects and related DOM nodes of the web page. VRAM is the dedicated memory of the GPU that manages data related to graphics; it is independent to the two previous measurements.

The figure shows that the used memory is much larger than the size of the 3D models. HEAP, which is more directly related to the X3D model, is approximately $8.5$ times the size of the model. This can be attributed to the following factors: (1) normal vectors and textures are not included in the X3D file (default of ViSL3D), they would double the size of the model if included; (2) The browser needs to convert the models to be applicable in the rendering engine. 

The RSS is larger since it includes HEAP and several other data and processes required for the rendering of the browser. Even so, the increase in RSS is lessened for large models, likely because the system begins using swap memory and cache. In addition, Figure \ref{fig:x3d-ram} highlights that rendering the models has a base RSS memory of $\sim 450$ MB, indicating that a large part of the used memory is not related to the model itself. This is confirmed by measuring the RSS of a blank or another less demanding web-page, which use at least $350$ MB. This property is also seen in VRAM, where the base memory is of $1$ GB. In this case, that VRAM could be reserved by X3DOM even if it is not completely used. However, it is evident that VRAM usage does not increase as rapidly as RSS or HEAP, suggesting that the GPU manages the model more efficiently.

The memory usage was analysed as part of the exploration described in Sec. \ref{available-tools}, revealing that other web-based visualisation applications were either comparable to X3DOM (e.g., VTK, glTF) or less efficient (e.g., Plotly). However, the amount of RAM required to render X3D models poses a challenge when dealing with very large datasets, occasionally leading to browser errors with larger models.
The first error (\textit{Invalid Array Length}) occurs due to the maximum size a JS array can have in web browser. This issue has been mitigated by splitting iso-surfaces into smaller parts when they exceed the array size limit, thereby splitting the resulting JS arrays. The second error (\textit{Out of Memory}) is triggered when the RSS RAM usage goes over $\approx 5000$MB; for this reason, Figure \ref{fig:x3d-ram} only shows models up to $\sim 400$ MB. These limitations are consistent across most common web browsers (Firefox, Chrome, Edge) as of 2024. However, they are expected to improve in the future, with increases in both the maximum array length and RAM limits.

This issue indicates that additional techniques are required to reduce the RAM usage and enable the visualisation of large astronomical datacubes. Some approaches are already implemented in ViSL3D; for instance, a preliminary low-resolution 3D model can be generated to provide an overview of the data, followed by high-resolution cutouts focused on specific features of interest. This method is particularly effective when creating the models on a remote server.

Other technologies that could optimise 3D rendering in web browsers include hierarchical 3D techniques (similar to HiPS), culling, binary compression methods (e.g., using .x3dbz or .glb formats), or algorithms designed to transfer more data into GPU memory. %However, these techniques require further development to fully realise their potential.
While the use of these technologies and techniques is undoubtedly important and relevant, it lies beyond the defined scope of this paper. Future work could explore this area in greater depth.

\subsection{Interactivity}

The approach we have developed offers a high degree of interactivity. It enables full customisation of 3D models, allowing users to select the number, values, and colours of the iso-surfaces. Users can examine the data from multiple perspectives by rotating and zooming the model, as well as hiding or showing individual iso-surfaces. This flexibility facilitates the detailed characterisation of extended features within the data.
The software also permits having various spectral lines from one or more datacubes in a single visualisation, allowing for a detailed comparison between lines. In addition, the option of including 2D images and markers at positions of galaxies from catalogues permits interactions with data from observations at different wavelengths. Markers can also be attached inside the model to highlight features in the data.
Further interactive options, described in Table \ref{tab_interactivity}, enhance the visualisation experience. All of these capabilities are accessible without requiring the installation of specialised software, thanks to the use of web technologies such as X3DOM and JS.

We have identified several improvements to the software that could enhance both its efficiency and interactive capabilities. One significant improvement is to adopt a CARTA-style hierarchical schema to provide histograms, spectra and other statistics from the dataset. Apart from creating the 3D model, this strategy would calculate statistics remotely and transfer them to the client. Another feature, which has been tested but not yet released, involves enabling the selection of coordinates by clicking directly on the 3D model. This functionality would allow for more precise determination of the positions of astrophysical features.

\section{Conclusions}\label{conclusion}

Next-generation observatories will produce an unprecedented volume of data. The SKAO is expected to generate approximately 700 PB/year, necessitating robust archiving solutions. This presents a substantial technical challenge across all stages of the scientific process and requires the SRCNet to function as a distributed and federated platform, serving as both data and service provider.

Among the essential services that the SRCNet must offer is data visualisation, which plays a critical role in the analysis of scientific data. While various tools are available for visualising data in 2D, some data products are inherently 3D, necessitating advanced 3D visualisation techniques. However, current solutions in this domain are not adequately equipped to meet the requirements of Big Data.
We have made an exploratory study of available software from astronomy and other fields for the visualisation of 3D data and used selected technologies to create a novel visualisation service.

We have built a prototype scientific archive within the espSRC that gives access to a visualisation service for 3D radio data, enabling the remote creation of 3D models and visualisation applications. The archive has been implemented using DaCHS and adheres to the SIA and Datalink IVOA recommendations to facilitate the discovery and provision of the service. Since there is no standard for server-side operations giving access to applications, similar to what SODA is for data access, we have presented two alternatives for this purpose: modifying SODA to include access to applications as well as data, or creating a new standard for discovery and access to sever side operations and processing, SOPA. We acknowledge that achieving consensus on this matter is complex; however, scientific platforms such as the SRCNet need a homogeneous way to describe and give access to operations, which can be defined using IVOA standards. The standardisation and integration into the archive enhance the service's accessibility while enabling remote server operations increases its scalability.

3D models are stored using the X3D standard and visualised through a web application powered by X3DOM, ensuring interoperability and a high degree of interactivity. The models represent iso-surfaces rather than full datacubes, effectively reducing their size while maintaining a faithful representation of the data. A custom web application is employed to render each model and enhance interactivity, offering features such as rotation, zooming, and the ability to hide or show individual surfaces, among others.

We have analysed the size and RAM usage of models generated from HI data of HCGs observed with the VLA and MeerKAT. The analysis indicates that at lower $S/N$ values, the size of the models increases, leading to higher RAM usage. This requires the use of cutouts and/or a lower resolution to visualise cubes larger than $400$ MB with various layers at $S/N\approx 3$. 
This limitation does not impact the effectiveness of visualisations for compact groups, as the most relevant information is typically confined to a small region of the datacube. However, it presents challenges for visualising larger structures at high resolutions. Further research is needed to explore methods for reducing memory usage and enhancing the capabilities of web-based visualisation software. Nevertheless, we consider this a subject for future investigation.

This work shows that a scientific archive can deliver advanced visualisation services for radio data, by generating 3D models remotely on a server, ensuring scalability to meet Big Data requirements while adhering to IVOA recommendations and FAIR principles. The proposed service offers enhanced capabilities for the interactive exploration of 3D data through a web application.

\section*{Acknowledgements}

The authors acknowledge financial support from the grant PID2021-123930OB-C21 funded by MICIU/AEI/10.13039/501100011033, by ERDF/EU and from the grant TED2021-130231B-I00 funded by MICIU/AEI and by the European Union NextGenerationEU/PRTR.

Ixaka Labadie-García acknowledges financial support from the grant PRE2021-100660 funded by MICIU/AEI and by ESF+.

María Ángeles Mendoza acknowledges financial support from the grant PTA2022-022026-I funded by MICIU/10.13039/501100011033 and by the FSE+

The authors acknowledge the Spanish Prototype of an SRC (SPSRC) service and support funded by the Ministerio de Ciencia, Innovación y Universidades (MICIU), by the Junta de Andalucía, by the European Regional Development Funds (ERDF) and by the European Union NextGenerationEU/PRTR. The SPSRC acknowledges financial support from the Agencia Estatal de Investigación (AEI) through the "Center of Excellence Severo Ochoa" award to the Instituto de Astrofísica de Andalucía (IAA-CSIC) (SEV-2017-0709) and from the grant CEX2021-001131-S funded by MICIU/AEI.

We acknowledge Bob Watson for inspiring the creation of the SOPA acronym.

\appendix

%\section{Appendix title 1}

% @INPROCEEDINGS{julian_srcnet2023,
%       author = {{Garrido}, Juli{\'a}n and {Verdes-Montenegro}, Lourdes and %{S{\'a}nchez}, Susana and {Gallardo}, Julio},
%        title = "{Status of the SKA project and the SKA Regional Centre %Network}",
%    booktitle = {Highlights on Spanish Astrophysics XI},
%         year = 2023,
%       month = may,
%        pages = {360},
%       adsurl = {https://ui.adsabs.harvard.edu/abs/2023hsa..conf..360G},
%      adsnote = {Provided by the SAO/NASA Astrophysics Data System}
%}

\bibliographystyle{elsarticle-harv} 
\bibliography{bibliography}

\begin{thebibliography}{57}
\expandafter\ifx\csname natexlab\endcsname\relax\def\natexlab#1{#1}\fi
\providecommand{\url}[1]{\texttt{#1}}
\providecommand{\href}[2]{#2}
\providecommand{\path}[1]{#1}
\providecommand{\DOIprefix}{doi:}
\providecommand{\ArXivprefix}{arXiv:}
\providecommand{\URLprefix}{URL: }
\providecommand{\Pubmedprefix}{pmid:}
\providecommand{\doi}[1]{\href{http://dx.doi.org/#1}{\path{#1}}}
\providecommand{\Pubmed}[1]{\href{pmid:#1}{\path{#1}}}
\providecommand{\bibinfo}[2]{#2}
\ifx\xfnm\relax \def\xfnm[#1]{\unskip,\space#1}\fi
%Type = Article
\bibitem[{Ahrens et~al.(2005)Ahrens, Geveci and Law}]{paraview2005}
\bibinfo{author}{Ahrens, J.}, \bibinfo{author}{Geveci, B.}, \bibinfo{author}{Law, C.}, \bibinfo{year}{2005}.
\newblock \bibinfo{title}{Paraview: An end-user tool for large data visualization}.
\newblock \bibinfo{journal}{Visualization Handbook} .
%Type = Inproceedings
\bibitem[{{Bacon} et~al.(2010){Bacon}, {Accardo}, {Adjali}, {Anwand}, {Bauer}, {Biswas}, {Blaizot}, {Boudon}, {Brau-Nogue}, {Brinchmann}, {Caillier}, {Capoani}, {Carollo}, {Contini}, {Couderc}, {Daguis{\'e}}, {Deiries}, {Delabre}, {Dreizler}, {Dubois}, {Dupieux}, {Dupuy}, {Emsellem}, {Fechner}, {Fleischmann}, {Fran{\c{c}}ois}, {Gallou}, {Gharsa}, {Glindemann}, {Gojak}, {Guiderdoni}, {Hansali}, {Hahn}, {Jarno}, {Kelz}, {Koehler}, {Kosmalski}, {Laurent}, {Le Floch}, {Lilly}, {Lizon}, {Loupias}, {Manescau}, {Monstein}, {Nicklas}, {Olaya}, {Pares}, {Pasquini}, {P{\'e}contal-Rousset}, {Pell{\'o}}, {Petit}, {Popow}, {Reiss}, {Remillieux}, {Renault}, {Roth}, {Rupprecht}, {Serre}, {Schaye}, {Soucail}, {Steinmetz}, {Streicher}, {Stuik}, {Valentin}, {Vernet}, {Weilbacher}, {Wisotzki} and {Yerle}}]{muse2010}
\bibinfo{author}{{Bacon}, R.}, \bibinfo{author}{{Accardo}, M.}, \bibinfo{author}{{Adjali}, L.}, \bibinfo{author}{{Anwand}, H.}, \bibinfo{author}{{Bauer}, S.}, \bibinfo{author}{{Biswas}, I.}, \bibinfo{author}{{Blaizot}, J.}, \bibinfo{author}{{Boudon}, D.}, \bibinfo{author}{{Brau-Nogue}, S.}, \bibinfo{author}{{Brinchmann}, J.}, \bibinfo{author}{{Caillier}, P.}, \bibinfo{author}{{Capoani}, L.}, \bibinfo{author}{{Carollo}, C.M.}, \bibinfo{author}{{Contini}, T.}, \bibinfo{author}{{Couderc}, P.}, \bibinfo{author}{{Daguis{\'e}}, E.}, \bibinfo{author}{{Deiries}, S.}, \bibinfo{author}{{Delabre}, B.}, \bibinfo{author}{{Dreizler}, S.}, \bibinfo{author}{{Dubois}, J.}, \bibinfo{author}{{Dupieux}, M.}, \bibinfo{author}{{Dupuy}, C.}, \bibinfo{author}{{Emsellem}, E.}, \bibinfo{author}{{Fechner}, T.}, \bibinfo{author}{{Fleischmann}, A.}, \bibinfo{author}{{Fran{\c{c}}ois}, M.}, \bibinfo{author}{{Gallou}, G.}, \bibinfo{author}{{Gharsa}, T.}, \bibinfo{author}{{Glindemann}, A.}, \bibinfo{author}{{Gojak}, D.},
  \bibinfo{author}{{Guiderdoni}, B.}, \bibinfo{author}{{Hansali}, G.}, \bibinfo{author}{{Hahn}, T.}, \bibinfo{author}{{Jarno}, A.}, \bibinfo{author}{{Kelz}, A.}, \bibinfo{author}{{Koehler}, C.}, \bibinfo{author}{{Kosmalski}, J.}, \bibinfo{author}{{Laurent}, F.}, \bibinfo{author}{{Le Floch}, M.}, \bibinfo{author}{{Lilly}, S.J.}, \bibinfo{author}{{Lizon}, J.L.}, \bibinfo{author}{{Loupias}, M.}, \bibinfo{author}{{Manescau}, A.}, \bibinfo{author}{{Monstein}, C.}, \bibinfo{author}{{Nicklas}, H.}, \bibinfo{author}{{Olaya}, J.C.}, \bibinfo{author}{{Pares}, L.}, \bibinfo{author}{{Pasquini}, L.}, \bibinfo{author}{{P{\'e}contal-Rousset}, A.}, \bibinfo{author}{{Pell{\'o}}, R.}, \bibinfo{author}{{Petit}, C.}, \bibinfo{author}{{Popow}, E.}, \bibinfo{author}{{Reiss}, R.}, \bibinfo{author}{{Remillieux}, A.}, \bibinfo{author}{{Renault}, E.}, \bibinfo{author}{{Roth}, M.}, \bibinfo{author}{{Rupprecht}, G.}, \bibinfo{author}{{Serre}, D.}, \bibinfo{author}{{Schaye}, J.}, \bibinfo{author}{{Soucail}, G.},
  \bibinfo{author}{{Steinmetz}, M.}, \bibinfo{author}{{Streicher}, O.}, \bibinfo{author}{{Stuik}, R.}, \bibinfo{author}{{Valentin}, H.}, \bibinfo{author}{{Vernet}, J.}, \bibinfo{author}{{Weilbacher}, P.}, \bibinfo{author}{{Wisotzki}, L.}, \bibinfo{author}{{Yerle}, N.}, \bibinfo{year}{2010}.
\newblock \bibinfo{title}{{The MUSE second-generation VLT instrument}}, in: \bibinfo{editor}{{McLean}, I.S.}, \bibinfo{editor}{{Ramsay}, S.K.}, \bibinfo{editor}{{Takami}, H.} (Eds.), \bibinfo{booktitle}{Ground-based and Airborne Instrumentation for Astronomy III}, p. \bibinfo{pages}{773508}.
\newblock \DOIprefix\doi{10.1117/12.856027}, \href{http://arxiv.org/abs/2211.16795}{{\tt arXiv:2211.16795}}.
%Type = Inproceedings
\bibitem[{{Barret} et~al.(2018){Barret}, {Lam Trong}, {den Herder}, {Piro}, {Cappi}, {Houvelin}, {Kelley}, {Mas-Hesse}, {Mitsuda}, {Paltani}, {Rauw}, {Rozanska}, {Wilms}, {Bandler}, {Barbera}, {Barcons}, {Bozzo}, {Ceballos}, {Charles}, {Costantini}, {Decourchelle}, {den Hartog}, {Duband}, {Duval}, {Fiore}, {Gatti}, {Goldwurm}, {Jackson}, {Jonker}, {Kilbourne}, {Macculi}, {Mendez}, {Molendi}, {Orleanski}, {Pajot}, {Pointecouteau}, {Porter}, {Pratt}, {Pr{\^e}le}, {Ravera}, {Sato}, {Schaye}, {Shinozaki}, {Thibert}, {Valenziano}, {Valette}, {Vink}, {Webb}, {Wise}, {Yamasaki}, {Douchin}, {Mesnager}, {Pontet}, {Pradines}, {Branduardi-Raymont}, {Bulbul}, {Dadina}, {Ettori}, {Finoguenov}, {Fukazawa}, {Janiuk}, {Kaastra}, {Mazzotta}, {Miller}, {Miniutti}, {Naze}, {Nicastro}, {Scioritino}, {Simonescu}, {Torrejon}, {Frezouls}, {Geoffray}, {Peille}, {Aicardi}, {Andr{\'e}}, {Daniel}, {Cl{\'e}net}, {Etcheverry}, {Gloaguen}, {Hervet}, {Jolly}, {Ledot}, {Paillet}, {Schmisser}, {Vella}, {Damery}, {Boyce}, {Dipirro}, {Lotti},
  {Schwander}, {Smith}, {Van Leeuwen}, {van Weers}, {Clerc}, {Cobo}, {Dauser}, {Kirsch}, {Cucchetti}, {Eckart}, {Ferrando} and {Natalucci}}]{athena_xifu2018}
\bibinfo{author}{{Barret}, D.}, \bibinfo{author}{{Lam Trong}, T.}, \bibinfo{author}{{den Herder}, J.W.}, \bibinfo{author}{{Piro}, L.}, \bibinfo{author}{{Cappi}, M.}, \bibinfo{author}{{Houvelin}, J.}, \bibinfo{author}{{Kelley}, R.}, \bibinfo{author}{{Mas-Hesse}, J.M.}, \bibinfo{author}{{Mitsuda}, K.}, \bibinfo{author}{{Paltani}, S.}, \bibinfo{author}{{Rauw}, G.}, \bibinfo{author}{{Rozanska}, A.}, \bibinfo{author}{{Wilms}, J.}, \bibinfo{author}{{Bandler}, S.}, \bibinfo{author}{{Barbera}, M.}, \bibinfo{author}{{Barcons}, X.}, \bibinfo{author}{{Bozzo}, E.}, \bibinfo{author}{{Ceballos}, M.T.}, \bibinfo{author}{{Charles}, I.}, \bibinfo{author}{{Costantini}, E.}, \bibinfo{author}{{Decourchelle}, A.}, \bibinfo{author}{{den Hartog}, R.}, \bibinfo{author}{{Duband}, L.}, \bibinfo{author}{{Duval}, J.M.}, \bibinfo{author}{{Fiore}, F.}, \bibinfo{author}{{Gatti}, F.}, \bibinfo{author}{{Goldwurm}, A.}, \bibinfo{author}{{Jackson}, B.}, \bibinfo{author}{{Jonker}, P.}, \bibinfo{author}{{Kilbourne}, C.},
  \bibinfo{author}{{Macculi}, C.}, \bibinfo{author}{{Mendez}, M.}, \bibinfo{author}{{Molendi}, S.}, \bibinfo{author}{{Orleanski}, P.}, \bibinfo{author}{{Pajot}, F.}, \bibinfo{author}{{Pointecouteau}, E.}, \bibinfo{author}{{Porter}, F.}, \bibinfo{author}{{Pratt}, G.W.}, \bibinfo{author}{{Pr{\^e}le}, D.}, \bibinfo{author}{{Ravera}, L.}, \bibinfo{author}{{Sato}, K.}, \bibinfo{author}{{Schaye}, J.}, \bibinfo{author}{{Shinozaki}, K.}, \bibinfo{author}{{Thibert}, T.}, \bibinfo{author}{{Valenziano}, L.}, \bibinfo{author}{{Valette}, V.}, \bibinfo{author}{{Vink}, J.}, \bibinfo{author}{{Webb}, N.}, \bibinfo{author}{{Wise}, M.}, \bibinfo{author}{{Yamasaki}, N.}, \bibinfo{author}{{Douchin}, F.}, \bibinfo{author}{{Mesnager}, J.M.}, \bibinfo{author}{{Pontet}, B.}, \bibinfo{author}{{Pradines}, A.}, \bibinfo{author}{{Branduardi-Raymont}, G.}, \bibinfo{author}{{Bulbul}, E.}, \bibinfo{author}{{Dadina}, M.}, \bibinfo{author}{{Ettori}, S.}, \bibinfo{author}{{Finoguenov}, A.}, \bibinfo{author}{{Fukazawa}, Y.},
  \bibinfo{author}{{Janiuk}, A.}, \bibinfo{author}{{Kaastra}, J.}, \bibinfo{author}{{Mazzotta}, P.}, \bibinfo{author}{{Miller}, J.}, \bibinfo{author}{{Miniutti}, G.}, \bibinfo{author}{{Naze}, Y.}, \bibinfo{author}{{Nicastro}, F.}, \bibinfo{author}{{Scioritino}, S.}, \bibinfo{author}{{Simonescu}, A.}, \bibinfo{author}{{Torrejon}, J.M.}, \bibinfo{author}{{Frezouls}, B.}, \bibinfo{author}{{Geoffray}, H.}, \bibinfo{author}{{Peille}, P.}, \bibinfo{author}{{Aicardi}, C.}, \bibinfo{author}{{Andr{\'e}}, J.}, \bibinfo{author}{{Daniel}, C.}, \bibinfo{author}{{Cl{\'e}net}, A.}, \bibinfo{author}{{Etcheverry}, C.}, \bibinfo{author}{{Gloaguen}, E.}, \bibinfo{author}{{Hervet}, G.}, \bibinfo{author}{{Jolly}, A.}, \bibinfo{author}{{Ledot}, A.}, \bibinfo{author}{{Paillet}, I.}, \bibinfo{author}{{Schmisser}, R.}, \bibinfo{author}{{Vella}, B.}, \bibinfo{author}{{Damery}, J.C.}, \bibinfo{author}{{Boyce}, K.}, \bibinfo{author}{{Dipirro}, M.}, \bibinfo{author}{{Lotti}, S.}, \bibinfo{author}{{Schwander}, D.},
  \bibinfo{author}{{Smith}, S.}, \bibinfo{author}{{Van Leeuwen}, B.J.}, \bibinfo{author}{{van Weers}, H.}, \bibinfo{author}{{Clerc}, N.}, \bibinfo{author}{{Cobo}, B.}, \bibinfo{author}{{Dauser}, T.}, \bibinfo{author}{{Kirsch}, C.}, \bibinfo{author}{{Cucchetti}, E.}, \bibinfo{author}{{Eckart}, M.}, \bibinfo{author}{{Ferrando}, P.}, \bibinfo{author}{{Natalucci}, L.}, \bibinfo{year}{2018}.
\newblock \bibinfo{title}{{The ATHENA X-ray Integral Field Unit (X-IFU)}}, in: \bibinfo{editor}{{den Herder}, J.W.A.}, \bibinfo{editor}{{Nikzad}, S.}, \bibinfo{editor}{{Nakazawa}, K.} (Eds.), \bibinfo{booktitle}{Space Telescopes and Instrumentation 2018: Ultraviolet to Gamma Ray}, p. \bibinfo{pages}{106991G}.
\newblock \DOIprefix\doi{10.1117/12.2312409}, \href{http://arxiv.org/abs/1807.06092}{{\tt arXiv:1807.06092}}.
%Type = Article
\bibitem[{{B{\"o}ker} et~al.(2022){B{\"o}ker}, {Arribas}, {L{\"u}tzgendorf}, {Alves de Oliveira}, {Beck}, {Birkmann}, {Bunker}, {Charlot}, {de Marchi}, {Ferruit}, {Giardino}, {Jakobsen}, {Kumari}, {L{\'o}pez-Caniego}, {Maiolino}, {Manjavacas}, {Marston}, {Moseley}, {Muzerolle}, {Ogle}, {Pirzkal}, {Rauscher}, {Rawle}, {Rix}, {Sabbi}, {Sargent}, {Sirianni}, {te Plate}, {Valenti}, {Willott} and {Zeidler}}]{nirspec2022}
\bibinfo{author}{{B{\"o}ker}, T.}, \bibinfo{author}{{Arribas}, S.}, \bibinfo{author}{{L{\"u}tzgendorf}, N.}, \bibinfo{author}{{Alves de Oliveira}, C.}, \bibinfo{author}{{Beck}, T.L.}, \bibinfo{author}{{Birkmann}, S.}, \bibinfo{author}{{Bunker}, A.J.}, \bibinfo{author}{{Charlot}, S.}, \bibinfo{author}{{de Marchi}, G.}, \bibinfo{author}{{Ferruit}, P.}, \bibinfo{author}{{Giardino}, G.}, \bibinfo{author}{{Jakobsen}, P.}, \bibinfo{author}{{Kumari}, N.}, \bibinfo{author}{{L{\'o}pez-Caniego}, M.}, \bibinfo{author}{{Maiolino}, R.}, \bibinfo{author}{{Manjavacas}, E.}, \bibinfo{author}{{Marston}, A.}, \bibinfo{author}{{Moseley}, S.H.}, \bibinfo{author}{{Muzerolle}, J.}, \bibinfo{author}{{Ogle}, P.}, \bibinfo{author}{{Pirzkal}, N.}, \bibinfo{author}{{Rauscher}, B.}, \bibinfo{author}{{Rawle}, T.}, \bibinfo{author}{{Rix}, H.W.}, \bibinfo{author}{{Sabbi}, E.}, \bibinfo{author}{{Sargent}, B.}, \bibinfo{author}{{Sirianni}, M.}, \bibinfo{author}{{te Plate}, M.}, \bibinfo{author}{{Valenti}, J.}, \bibinfo{author}{{Willott},
  C.J.}, \bibinfo{author}{{Zeidler}, P.}, \bibinfo{year}{2022}.
\newblock \bibinfo{title}{{The Near-Infrared Spectrograph (NIRSpec) on the James Webb Space Telescope. III. Integral-field spectroscopy}}.
\newblock \bibinfo{journal}{\aap} \bibinfo{volume}{661}, \bibinfo{pages}{A82}.
\newblock \DOIprefix\doi{10.1051/0004-6361/202142589}, \href{http://arxiv.org/abs/2202.03308}{{\tt arXiv:2202.03308}}.
%Type = Misc
\bibitem[{{Bolton} et~al.(2023){Bolton}, {Breen}, A. and {SKAO Regional Centre Steering Committee}}]{bolton_srcnet2023}
\bibinfo{author}{{Bolton}, R.}, \bibinfo{author}{{Breen}, S.}, \bibinfo{author}{A., C.}, \bibinfo{author}{{SKAO Regional Centre Steering Committee}}, \bibinfo{year}{2023}.
\newblock \bibinfo{title}{Src network vision and principles}.
\newblock \URLprefix \url{https://indico.skatelescope.org/category/271/attachments/9795/17084/SRC-0000005_01_SRC%20Network%20Vision%20and%20Principles%20-%20signed.pdf}.
%Type = Misc
\bibitem[{{Bonnarel} et~al.(2017){Bonnarel}, {Dowler}, {Demleitner}, {Tody} and {Dempsey}}]{soda1.0}
\bibinfo{author}{{Bonnarel}, F.}, \bibinfo{author}{{Dowler}, P.}, \bibinfo{author}{{Demleitner}, M.}, \bibinfo{author}{{Tody}, D.}, \bibinfo{author}{{Dempsey}, J.}, \bibinfo{year}{2017}.
\newblock \bibinfo{title}{{IVOA Server-side Operations for Data Access Version 1.0}}.
\newblock \bibinfo{howpublished}{IVOA Recommendation 17 May 2017}.
\newblock \DOIprefix\doi{10.5479/ADS/bib/2017ivoa.spec.0517B}, \href{http://arxiv.org/abs/1710.08791}{{\tt arXiv:1710.08791}}.
%Type = Misc
\bibitem[{{Bonnarel} et~al.(2023){Bonnarel}, {Dowler}, {Michel}, {Demleitner} and {Taylor}}]{datalink1.1}
\bibinfo{author}{{Bonnarel}, F.}, \bibinfo{author}{{Dowler}, P.}, \bibinfo{author}{{Michel}, L.}, \bibinfo{author}{{Demleitner}, M.}, \bibinfo{author}{{Taylor}, M.}, \bibinfo{year}{2023}.
\newblock \bibinfo{title}{{IVOA DataLink Version 1.1}}.
\newblock \bibinfo{howpublished}{IVOA Recommendation 15 December 2023}.
\newblock \DOIprefix\doi{10.5479/ADS/bib/2023ivoa.spec.1215B}.
%Type = Article
\bibitem[{{Bonnarel} et~al.(2000){Bonnarel}, {Fernique}, {Bienaym{\'e}}, {Egret}, {Genova}, {Louys}, {Ochsenbein}, {Wenger} and {Bartlett}}]{Aladin2000}
\bibinfo{author}{{Bonnarel}, F.}, \bibinfo{author}{{Fernique}, P.}, \bibinfo{author}{{Bienaym{\'e}}, O.}, \bibinfo{author}{{Egret}, D.}, \bibinfo{author}{{Genova}, F.}, \bibinfo{author}{{Louys}, M.}, \bibinfo{author}{{Ochsenbein}, F.}, \bibinfo{author}{{Wenger}, M.}, \bibinfo{author}{{Bartlett}, J.G.}, \bibinfo{year}{2000}.
\newblock \bibinfo{title}{{The ALADIN interactive sky atlas. A reference tool for identification of astronomical sources}}.
\newblock \bibinfo{journal}{\aaps} \bibinfo{volume}{143}, \bibinfo{pages}{33--40}.
\newblock \DOIprefix\doi{10.1051/aas:2000331}.
%Type = Article
\bibitem[{Bouyssié et~al.(2019)Bouyssié, Lesne, Locard-Paulet, Albigot, Burlet-Schiltz and Marcoux}]{3Dprotein2019}
\bibinfo{author}{Bouyssié, D.}, \bibinfo{author}{Lesne, J.}, \bibinfo{author}{Locard-Paulet, M.}, \bibinfo{author}{Albigot, R.}, \bibinfo{author}{Burlet-Schiltz, O.}, \bibinfo{author}{Marcoux, J.}, \bibinfo{year}{2019}.
\newblock \bibinfo{title}{Hdx-viewer: interactive 3d visualization of hydrogen-deuterium exchange data}.
\newblock \bibinfo{journal}{Bioinformatics (Oxford, England)} \bibinfo{volume}{35}.
\newblock \DOIprefix\doi{10.1093/bioinformatics/btz550}.
%Type = Misc
\bibitem[{Chue~Hong et~al.(2022)Chue~Hong, Katz, Barker, Lamprecht, Martinez, Psomopoulos, Harrow, Castro, Gruenpeter, Martinez, Honeyman, Struck, Lee, Loewe, van Werkhoven, Jones, Garijo, Plomp, Genova, Shanahan, Leng, Hellström, Sandström, Sinha, Kuzak, Herterich, Zhang, Islam, Sansone, Pollard, Atmojo, Williams, Czerniak, Niehues, Fouilloux, Desinghu, Goble, Richard, Gray, Erdmann, Nüst, Tartarini, Ranguelova, Anzt, Todorov, McNally, Moldon, Burnett, Garrido-Sánchez, Belhajjame, Sesink, Hwang, Tovani-Palone, Wilkinson, Servillat, Liffers, Fox, Miljković, Lynch, Martinez~Lavanchy, Gesing, Stevens, Martinez~Cuesta, Peroni, Soiland-Reyes, Bakker, Rabemanantsoa, Sochat, Yehudi and WG}]{FAIR_Software2022}
\bibinfo{author}{Chue~Hong, N.P.}, \bibinfo{author}{Katz, D.S.}, \bibinfo{author}{Barker, M.}, \bibinfo{author}{Lamprecht, A.L.}, \bibinfo{author}{Martinez, C.}, \bibinfo{author}{Psomopoulos, F.E.}, \bibinfo{author}{Harrow, J.}, \bibinfo{author}{Castro, L.J.}, \bibinfo{author}{Gruenpeter, M.}, \bibinfo{author}{Martinez, P.A.}, \bibinfo{author}{Honeyman, T.}, \bibinfo{author}{Struck, A.}, \bibinfo{author}{Lee, A.}, \bibinfo{author}{Loewe, A.}, \bibinfo{author}{van Werkhoven, B.}, \bibinfo{author}{Jones, C.}, \bibinfo{author}{Garijo, D.}, \bibinfo{author}{Plomp, E.}, \bibinfo{author}{Genova, F.}, \bibinfo{author}{Shanahan, H.}, \bibinfo{author}{Leng, J.}, \bibinfo{author}{Hellström, M.}, \bibinfo{author}{Sandström, M.}, \bibinfo{author}{Sinha, M.}, \bibinfo{author}{Kuzak, M.}, \bibinfo{author}{Herterich, P.}, \bibinfo{author}{Zhang, Q.}, \bibinfo{author}{Islam, S.}, \bibinfo{author}{Sansone, S.A.}, \bibinfo{author}{Pollard, T.}, \bibinfo{author}{Atmojo, U.D.}, \bibinfo{author}{Williams, A.},
  \bibinfo{author}{Czerniak, A.}, \bibinfo{author}{Niehues, A.}, \bibinfo{author}{Fouilloux, A.C.}, \bibinfo{author}{Desinghu, B.}, \bibinfo{author}{Goble, C.}, \bibinfo{author}{Richard, C.}, \bibinfo{author}{Gray, C.}, \bibinfo{author}{Erdmann, C.}, \bibinfo{author}{Nüst, D.}, \bibinfo{author}{Tartarini, D.}, \bibinfo{author}{Ranguelova, E.}, \bibinfo{author}{Anzt, H.}, \bibinfo{author}{Todorov, I.}, \bibinfo{author}{McNally, J.}, \bibinfo{author}{Moldon, J.}, \bibinfo{author}{Burnett, J.}, \bibinfo{author}{Garrido-Sánchez, J.}, \bibinfo{author}{Belhajjame, K.}, \bibinfo{author}{Sesink, L.}, \bibinfo{author}{Hwang, L.}, \bibinfo{author}{Tovani-Palone, M.R.}, \bibinfo{author}{Wilkinson, M.D.}, \bibinfo{author}{Servillat, M.}, \bibinfo{author}{Liffers, M.}, \bibinfo{author}{Fox, M.}, \bibinfo{author}{Miljković, N.}, \bibinfo{author}{Lynch, N.}, \bibinfo{author}{Martinez~Lavanchy, P.}, \bibinfo{author}{Gesing, S.}, \bibinfo{author}{Stevens, S.}, \bibinfo{author}{Martinez~Cuesta, S.}, \bibinfo{author}{Peroni,
  S.}, \bibinfo{author}{Soiland-Reyes, S.}, \bibinfo{author}{Bakker, T.}, \bibinfo{author}{Rabemanantsoa, T.}, \bibinfo{author}{Sochat, V.}, \bibinfo{author}{Yehudi, Y.}, \bibinfo{author}{WG, R.F.}, \bibinfo{year}{2022}.
\newblock \bibinfo{title}{{FAIR Principles for Research Software (FAIR4RS Principles)}}.
\newblock \URLprefix \url{https://doi.org/10.15497/RDA00068}, \DOIprefix\doi{10.15497/RDA00068}.
%Type = Article
\bibitem[{{Comrie} et~al.(2020){Comrie}, {Pi{\'n}ska}, {Simmonds} and {Taylor}}]{carta2020}
\bibinfo{author}{{Comrie}, A.}, \bibinfo{author}{{Pi{\'n}ska}, A.}, \bibinfo{author}{{Simmonds}, R.}, \bibinfo{author}{{Taylor}, A.R.}, \bibinfo{year}{2020}.
\newblock \bibinfo{title}{{Development and application of an HDF5 schema for SKA-scale image cube visualization}}.
\newblock \bibinfo{journal}{Astronomy and Computing} \bibinfo{volume}{32}, \bibinfo{pages}{100389}.
\newblock \DOIprefix\doi{10.1016/j.ascom.2020.100389}.
%Type = Misc
\bibitem[{Comrie et~al.(2021)Comrie, Wang, Hsu, Moraghan, Harris, Pang, Pińska, Chiang, Chang, Hwang, Jan, Lin and Simmonds}]{carta_software}
\bibinfo{author}{Comrie, A.}, \bibinfo{author}{Wang, K.S.}, \bibinfo{author}{Hsu, S.C.}, \bibinfo{author}{Moraghan, A.}, \bibinfo{author}{Harris, P.}, \bibinfo{author}{Pang, Q.}, \bibinfo{author}{Pińska, A.}, \bibinfo{author}{Chiang, C.C.}, \bibinfo{author}{Chang, T.H.}, \bibinfo{author}{Hwang, Y.H.}, \bibinfo{author}{Jan, H.}, \bibinfo{author}{Lin, M.Y.}, \bibinfo{author}{Simmonds, R.}, \bibinfo{year}{2021}.
\newblock \bibinfo{title}{{CARTA: The Cube Analysis and Rendering Tool for Astronomy}}.
\newblock \URLprefix \url{https://doi.org/10.5281/zenodo.4905459}, \DOIprefix\doi{10.5281/zenodo.4905459}.
%Type = Misc
\bibitem[{{Demleitner} and {Harrison}(2019)}]{useofcapabilitiesVO}
\bibinfo{author}{{Demleitner}, M.}, \bibinfo{author}{{Harrison}, P.}, \bibinfo{year}{2019}.
\newblock \bibinfo{title}{{On the Use of Capabilities in the VO Version 1.0}}.
\newblock \bibinfo{howpublished}{IVOA Note 2019-03-15}.
\newblock \URLprefix \url{http://www.ivoa.net/documents/caproles/20190315}.
%Type = Article
\bibitem[{{Demleitner} et~al.(2014){Demleitner}, {Neves}, {Rothmaier} and {Wambsganss}}]{dachs2014}
\bibinfo{author}{{Demleitner}, M.}, \bibinfo{author}{{Neves}, M.C.}, \bibinfo{author}{{Rothmaier}, F.}, \bibinfo{author}{{Wambsganss}, J.}, \bibinfo{year}{2014}.
\newblock \bibinfo{title}{{Virtual observatory publishing with DaCHS}}.
\newblock \bibinfo{journal}{Astronomy and Computing} \bibinfo{volume}{7}, \bibinfo{pages}{27--36}.
\newblock \DOIprefix\doi{10.1016/j.ascom.2014.08.003}, \href{http://arxiv.org/abs/1408.5733}{{\tt arXiv:1408.5733}}.
%Type = Misc
\bibitem[{{Dower} et~al.(2018){Dower}, {Demleitner}, {Benson}, {Plante}, {Auden}, {Graham}, {Greene}, {Hill}, {Linde}, {Morris}, {O`Mullane}, {Rixon}, {St{\'e}b{\'e}} and {Andrews}}]{Registry1.1}
\bibinfo{author}{{Dower}, T.}, \bibinfo{author}{{Demleitner}, M.}, \bibinfo{author}{{Benson}, K.}, \bibinfo{author}{{Plante}, R.}, \bibinfo{author}{{Auden}, E.}, \bibinfo{author}{{Graham}, M.}, \bibinfo{author}{{Greene}, G.}, \bibinfo{author}{{Hill}, M.}, \bibinfo{author}{{Linde}, T.}, \bibinfo{author}{{Morris}, D.}, \bibinfo{author}{{O`Mullane}, W.}, \bibinfo{author}{{Rixon}, G.}, \bibinfo{author}{{St{\'e}b{\'e}}, A.}, \bibinfo{author}{{Andrews}, K.}, \bibinfo{year}{2018}.
\newblock \bibinfo{title}{{Registry Interfaces Version 1.1}}.
\newblock \bibinfo{howpublished}{IVOA Recommendation 23 July 2018}.
\newblock \DOIprefix\doi{10.5479/ADS/bib/2018ivoa.spec.0723D}.
%Type = Misc
\bibitem[{{Dowler} et~al.(2015){Dowler}, {Bonnarel} and {Tody}}]{sia2.0}
\bibinfo{author}{{Dowler}, P.}, \bibinfo{author}{{Bonnarel}, F.}, \bibinfo{author}{{Tody}, D.}, \bibinfo{year}{2015}.
\newblock \bibinfo{title}{{IVOA Simple Image Access Version 2.0}}.
\newblock \bibinfo{howpublished}{IVOA Recommendation 23 December 2015}.
\newblock \DOIprefix\doi{10.5479/ADS/bib/2015ivoa.spec.1223D}.
%Type = Misc
\bibitem[{{Dowler} et~al.(2017){Dowler}, {Demleitner}, {Taylor} and {Tody}}]{DALI1.1}
\bibinfo{author}{{Dowler}, P.}, \bibinfo{author}{{Demleitner}, M.}, \bibinfo{author}{{Taylor}, M.}, \bibinfo{author}{{Tody}, D.}, \bibinfo{year}{2017}.
\newblock \bibinfo{title}{{Data Access Layer Interface Version 1.1}}.
\newblock \bibinfo{howpublished}{IVOA Recommendation 17 May 2017}.
\newblock \DOIprefix\doi{10.5479/ADS/bib/2017ivoa.spec.0517D}.
%Type = Misc
\bibitem[{{Dowler} et~al.(2019){Dowler}, {Rixon}, {Tody} and {Demleitner}}]{tap1.1}
\bibinfo{author}{{Dowler}, P.}, \bibinfo{author}{{Rixon}, G.}, \bibinfo{author}{{Tody}, D.}, \bibinfo{author}{{Demleitner}, M.}, \bibinfo{year}{2019}.
\newblock \bibinfo{title}{{Table Access Protocol Version 1.1}}.
\newblock \bibinfo{howpublished}{IVOA Recommendation 27 September 2019}.
\newblock \DOIprefix\doi{10.5479/ADS/bib/2019ivoa.spec.0927D}.
%Type = Misc
\bibitem[{{Fernique}(2024)}]{hipsgen3d}
\bibinfo{author}{{Fernique}, P.}, \bibinfo{year}{2024}.
\newblock \bibinfo{title}{{Hipsgen: news features for news HiPS}}.
%Type = Misc
\bibitem[{{Fernique} et~al.(2017){Fernique}, {Allen}, {Boch}, {Donaldson}, {Durand}, {Ebisawa}, {Michel}, {Salgado} and {Stoehr}}]{IVOA_HIPS2017}
\bibinfo{author}{{Fernique}, P.}, \bibinfo{author}{{Allen}, M.}, \bibinfo{author}{{Boch}, T.}, \bibinfo{author}{{Donaldson}, T.}, \bibinfo{author}{{Durand}, D.}, \bibinfo{author}{{Ebisawa}, K.}, \bibinfo{author}{{Michel}, L.}, \bibinfo{author}{{Salgado}, J.}, \bibinfo{author}{{Stoehr}, F.}, \bibinfo{year}{2017}.
\newblock \bibinfo{title}{{HiPS - Hierarchical Progressive Survey Version 1.0}}.
\newblock \bibinfo{howpublished}{IVOA Recommendation 19 May 2017}.
\newblock \DOIprefix\doi{10.5479/ADS/bib/2017ivoa.spec.0519F}, \href{http://arxiv.org/abs/1708.09704}{{\tt arXiv:1708.09704}}.
%Type = Article
\bibitem[{{Fernique} et~al.(2015){Fernique}, {Allen}, {Boch}, {Oberto}, {Pineau}, {Durand}, {Bot}, {Cambr{\'e}sy}, {Derriere}, {Genova} and {Bonnarel}}]{hips2015}
\bibinfo{author}{{Fernique}, P.}, \bibinfo{author}{{Allen}, M.G.}, \bibinfo{author}{{Boch}, T.}, \bibinfo{author}{{Oberto}, A.}, \bibinfo{author}{{Pineau}, F.X.}, \bibinfo{author}{{Durand}, D.}, \bibinfo{author}{{Bot}, C.}, \bibinfo{author}{{Cambr{\'e}sy}, L.}, \bibinfo{author}{{Derriere}, S.}, \bibinfo{author}{{Genova}, F.}, \bibinfo{author}{{Bonnarel}, F.}, \bibinfo{year}{2015}.
\newblock \bibinfo{title}{{Hierarchical progressive surveys. Multi-resolution HEALPix data structures for astronomical images, catalogues, and 3-dimensional data cubes}}.
\newblock \bibinfo{journal}{\aap} \bibinfo{volume}{578}, \bibinfo{pages}{A114}.
\newblock \DOIprefix\doi{10.1051/0004-6361/201526075}, \href{http://arxiv.org/abs/1505.02291}{{\tt arXiv:1505.02291}}.
%Type = Article
\bibitem[{{Garrido} et~al.(2022){Garrido}, {Darriba}, {S{\'a}nchez-Exp{\'o}sito}, {Parra-Roy{\'o}n}, {Mold{\'o}n}, {Mendoza}, {Luna-Valero}, {Alberdi}, {M{\'a}rquez} and {Verdes-Montenegro}}]{espsrc2022}
\bibinfo{author}{{Garrido}, J.}, \bibinfo{author}{{Darriba}, L.}, \bibinfo{author}{{S{\'a}nchez-Exp{\'o}sito}, S.}, \bibinfo{author}{{Parra-Roy{\'o}n}, M.}, \bibinfo{author}{{Mold{\'o}n}, J.}, \bibinfo{author}{{Mendoza}, M.{\'A}.}, \bibinfo{author}{{Luna-Valero}, S.}, \bibinfo{author}{{Alberdi}, A.}, \bibinfo{author}{{M{\'a}rquez}, I.}, \bibinfo{author}{{Verdes-Montenegro}, L.}, \bibinfo{year}{2022}.
\newblock \bibinfo{title}{{Toward a Spanish SKA Regional Centre fully engaged with open science}}.
\newblock \bibinfo{journal}{Journal of Astronomical Telescopes, Instruments, and Systems} \bibinfo{volume}{8}, \bibinfo{pages}{011004}.
\newblock \DOIprefix\doi{10.1117/1.JATIS.8.1.011004}.
%Type = Inproceedings
\bibitem[{{Gaudet} et~al.(2010){Gaudet}, {Hill}, {Armstrong}, {Ball}, {Burke}, {Chapel}, {Chapin}, {Damian}, {Dowler}, {Gable}, {Goliath}, {Ghiurea}, {Fabbro}, {Gwyn}, {Jenkins}, {Kavelaars}, {Major}, {Ouellette}, {Paterson}, {Peddle}, {Penfold-Brown}, {Pritchet}, {Schade}, {Sobie}, {Woods}, {Yeung} and {Zhang}}]{CANFAR}
\bibinfo{author}{{Gaudet}, S.}, \bibinfo{author}{{Hill}, N.}, \bibinfo{author}{{Armstrong}, P.}, \bibinfo{author}{{Ball}, N.}, \bibinfo{author}{{Burke}, J.}, \bibinfo{author}{{Chapel}, B.}, \bibinfo{author}{{Chapin}, E.}, \bibinfo{author}{{Damian}, A.}, \bibinfo{author}{{Dowler}, P.}, \bibinfo{author}{{Gable}, I.}, \bibinfo{author}{{Goliath}, S.}, \bibinfo{author}{{Ghiurea}, I.}, \bibinfo{author}{{Fabbro}, S.}, \bibinfo{author}{{Gwyn}, S.}, \bibinfo{author}{{Jenkins}, D.}, \bibinfo{author}{{Kavelaars}, J.}, \bibinfo{author}{{Major}, B.}, \bibinfo{author}{{Ouellette}, J.}, \bibinfo{author}{{Paterson}, M.}, \bibinfo{author}{{Peddle}, M.}, \bibinfo{author}{{Penfold-Brown}, D.}, \bibinfo{author}{{Pritchet}, C.}, \bibinfo{author}{{Schade}, D.}, \bibinfo{author}{{Sobie}, R.}, \bibinfo{author}{{Woods}, D.}, \bibinfo{author}{{Yeung}, A.}, \bibinfo{author}{{Zhang}, Y.}, \bibinfo{year}{2010}.
\newblock \bibinfo{title}{{CANFAR: the Canadian Advanced Network for Astronomical Research}}, in: \bibinfo{editor}{{Radziwill}, N.M.}, \bibinfo{editor}{{Bridger}, A.} (Eds.), \bibinfo{booktitle}{Software and Cyberinfrastructure for Astronomy}, p. \bibinfo{pages}{77401I}.
\newblock \DOIprefix\doi{10.1117/12.858026}.
%Type = Misc
\bibitem[{{Graham} et~al.(2017){Graham}, {Rixon}, {Dowler}, {Major}, {Grid} and {Web Services Working Group}}]{VOSI1.1}
\bibinfo{author}{{Graham}, M.}, \bibinfo{author}{{Rixon}, G.}, \bibinfo{author}{{Dowler}, P.}, \bibinfo{author}{{Major}, B.}, \bibinfo{author}{{Grid}}, \bibinfo{author}{{Web Services Working Group}}, \bibinfo{year}{2017}.
\newblock \bibinfo{title}{{IVOA Support Interfaces Version 1.1}}.
\newblock \bibinfo{howpublished}{IVOA Recommendation 24 May 2017}.
\newblock \DOIprefix\doi{10.5479/ADS/bib/2017ivoa.spec.0524G}.
%Type = Article
\bibitem[{Hassan and Fluke(2011)}]{hassanfluke2011Petascale}
\bibinfo{author}{Hassan, A.}, \bibinfo{author}{Fluke, C.J.}, \bibinfo{year}{2011}.
\newblock \bibinfo{title}{Scientific visualization in astronomy: Towards the petascale astronomy era}.
\newblock \bibinfo{journal}{Publications of the Astronomical Society of Australia} \bibinfo{volume}{28}, \bibinfo{pages}{150–170}.
\newblock \DOIprefix\doi{10.1071/AS10031}.
%Type = Article
\bibitem[{He et~al.(2023)He, Liu, He and Cao}]{metaverse2023}
\bibinfo{author}{He, L.}, \bibinfo{author}{Liu, K.}, \bibinfo{author}{He, Z.}, \bibinfo{author}{Cao, L.}, \bibinfo{year}{2023}.
\newblock \bibinfo{title}{Three-dimensional holographic communication system for the metaverse}.
\newblock \bibinfo{journal}{Optics Communications} \bibinfo{volume}{526}, \bibinfo{pages}{128894}.
\newblock \URLprefix \url{https://www.sciencedirect.com/science/article/pii/S0030401822005661}, \DOIprefix\doi{https://doi.org/10.1016/j.optcom.2022.128894}.
%Type = Article
\bibitem[{{Hickson}(1982)}]{Hickson1982}
\bibinfo{author}{{Hickson}, P.}, \bibinfo{year}{1982}.
\newblock \bibinfo{title}{{Systematic properties of compact groups of galaxies.}}
\newblock \bibinfo{journal}{\apj} \bibinfo{volume}{255}, \bibinfo{pages}{382--391}.
\newblock \DOIprefix\doi{10.1086/159838}.
%Type = Inproceedings
\bibitem[{{Jarrett} et~al.(2024){Jarrett}, {Comrie}, {Sivitilli}, {Pretorius}, {Vitello} and {Marchetti}}]{idavie2024}
\bibinfo{author}{{Jarrett}, T.}, \bibinfo{author}{{Comrie}, A.}, \bibinfo{author}{{Sivitilli}, A.}, \bibinfo{author}{{Pretorius}, P.C.}, \bibinfo{author}{{Vitello}, F.}, \bibinfo{author}{{Marchetti}, L.}, \bibinfo{year}{2024}.
\newblock \bibinfo{title}{{iDaVIE: Immersive Data Visualisation Interactive Explorer}}, in: \bibinfo{booktitle}{Zenodo Software}, \bibinfo{publisher}{Zenodo}. p. \bibinfo{pages}{4614115}.
\newblock \DOIprefix\doi{10.5281/zenodo.4614115}.
%Type = Inproceedings
\bibitem[{{Jonas} and {MeerKAT Team}(2016)}]{meerkat2016}
\bibinfo{author}{{Jonas}, J.}, \bibinfo{author}{{MeerKAT Team}}, \bibinfo{year}{2016}.
\newblock \bibinfo{title}{{The MeerKAT Radio Telescope}}, in: \bibinfo{booktitle}{MeerKAT Science: On the Pathway to the SKA}, p.~\bibinfo{pages}{1}.
\newblock \DOIprefix\doi{10.22323/1.277.0001}.
%Type = Article
\bibitem[{{Jones} et~al.(2019){Jones}, {Verdes-Montenegro}, {Damas-Segovia}, {Borthakur}, {Yun}, {del Olmo}, {Perea}, {Rom{\'a}n}, {Luna}, {Lopez Gutierrez}, {Williams}, {Vogt}, {Garrido}, {Sanchez}, {Cannon} and {Ram{\'\i}rez-Moreta}}]{Jones2019hcg16}
\bibinfo{author}{{Jones}, M.G.}, \bibinfo{author}{{Verdes-Montenegro}, L.}, \bibinfo{author}{{Damas-Segovia}, A.}, \bibinfo{author}{{Borthakur}, S.}, \bibinfo{author}{{Yun}, M.}, \bibinfo{author}{{del Olmo}, A.}, \bibinfo{author}{{Perea}, J.}, \bibinfo{author}{{Rom{\'a}n}, J.}, \bibinfo{author}{{Luna}, S.}, \bibinfo{author}{{Lopez Gutierrez}, D.}, \bibinfo{author}{{Williams}, B.}, \bibinfo{author}{{Vogt}, F.P.A.}, \bibinfo{author}{{Garrido}, J.}, \bibinfo{author}{{Sanchez}, S.}, \bibinfo{author}{{Cannon}, J.}, \bibinfo{author}{{Ram{\'\i}rez-Moreta}, P.}, \bibinfo{year}{2019}.
\newblock \bibinfo{title}{{Evolution of compact groups from intermediate to final stages. A case study of the H I content of HCG 16}}.
\newblock \bibinfo{journal}{\aap} \bibinfo{volume}{632}, \bibinfo{pages}{A78}.
\newblock \DOIprefix\doi{10.1051/0004-6361/201936349}, \href{http://arxiv.org/abs/1910.03420}{{\tt arXiv:1910.03420}}.
%Type = Article
\bibitem[{{Jones} et~al.(2023){Jones}, {Verdes-Montenegro}, {Moldon}, {Damas Segovia}, {Borthakur}, {Luna}, {Yun}, {del Olmo}, {Perea}, {Cannon}, {Lopez Gutierrez}, {Cluver}, {Garrido} and {Sanchez}}]{VLAHCG}
\bibinfo{author}{{Jones}, M.G.}, \bibinfo{author}{{Verdes-Montenegro}, L.}, \bibinfo{author}{{Moldon}, J.}, \bibinfo{author}{{Damas Segovia}, A.}, \bibinfo{author}{{Borthakur}, S.}, \bibinfo{author}{{Luna}, S.}, \bibinfo{author}{{Yun}, M.}, \bibinfo{author}{{del Olmo}, A.}, \bibinfo{author}{{Perea}, J.}, \bibinfo{author}{{Cannon}, J.}, \bibinfo{author}{{Lopez Gutierrez}, D.}, \bibinfo{author}{{Cluver}, M.}, \bibinfo{author}{{Garrido}, J.}, \bibinfo{author}{{Sanchez}, S.}, \bibinfo{year}{2023}.
\newblock \bibinfo{title}{{Disturbed, diffuse, or just missing? A global study of the H I content of Hickson compact groups}}.
\newblock \bibinfo{journal}{\aap} \bibinfo{volume}{670}, \bibinfo{pages}{A21}.
\newblock \DOIprefix\doi{10.1051/0004-6361/202244622}, \href{http://arxiv.org/abs/2211.15687}{{\tt arXiv:2211.15687}}.
%Type = Misc
\bibitem[{Labadie-García et~al.(2024)Labadie-García, Garrido and Verdes-Montenegro}]{visl3d}
\bibinfo{author}{Labadie-García, I.}, \bibinfo{author}{Garrido, J.}, \bibinfo{author}{Verdes-Montenegro, L.}, \bibinfo{year}{2024}.
\newblock \bibinfo{title}{ixakalabadie/visl3d: v0.2}.
\newblock \URLprefix \url{https://doi.org/10.5281/zenodo.13254756}, \DOIprefix\doi{10.5281/zenodo.13254756}.
%Type = Article
\bibitem[{{Lan} et~al.(2021){Lan}, {Young}, {Anderson}, {Ynnerman}, {Bock}, {Borkin}, {Forbes}, {Kollmeier} and {Wang}}]{astrovis2021}
\bibinfo{author}{{Lan}, F.}, \bibinfo{author}{{Young}, M.}, \bibinfo{author}{{Anderson}, L.}, \bibinfo{author}{{Ynnerman}, A.}, \bibinfo{author}{{Bock}, A.}, \bibinfo{author}{{Borkin}, M.A.}, \bibinfo{author}{{Forbes}, A.G.}, \bibinfo{author}{{Kollmeier}, J.A.}, \bibinfo{author}{{Wang}, B.}, \bibinfo{year}{2021}.
\newblock \bibinfo{title}{{Visualization in Astrophysics: Developing New Methods, Discovering Our Universe, and Educating the Earth}}.
\newblock \bibinfo{journal}{arXiv e-prints} , \bibinfo{pages}{arXiv:2106.00152}\DOIprefix\doi{10.48550/arXiv.2106.00152}, \href{http://arxiv.org/abs/2106.00152}{{\tt arXiv:2106.00152}}.
%Type = Article
\bibitem[{{Lesne} et~al.(2020){Lesne}, {Locard-Paulet}, {Parra}, {Zivkovi{\'c}}, {Menneteau}, {Bousquet}, {Burlet-Schiltz} and {Marcoux}}]{protein2020}
\bibinfo{author}{{Lesne}, J.}, \bibinfo{author}{{Locard-Paulet}, M.}, \bibinfo{author}{{Parra}, J.}, \bibinfo{author}{{Zivkovi{\'c}}, D.}, \bibinfo{author}{{Menneteau}, T.}, \bibinfo{author}{{Bousquet}, M.P.}, \bibinfo{author}{{Burlet-Schiltz}, O.}, \bibinfo{author}{{Marcoux}, J.}, \bibinfo{year}{2020}.
\newblock \bibinfo{title}{{Conformational maps of human 20S proteasomes reveal PA28- and immuno-dependent inter-ring crosstalks}}.
\newblock \bibinfo{journal}{Nature Communications} \bibinfo{volume}{11}, \bibinfo{pages}{6140}.
\newblock \DOIprefix\doi{10.1038/s41467-020-19934-z}.
%Type = Article
\bibitem[{{Lewiner} et~al.(2003){Lewiner}, {Lopes}, {Wilson Vieira} and {Tavares}}]{marching2003}
\bibinfo{author}{{Lewiner}, T.}, \bibinfo{author}{{Lopes}, H.}, \bibinfo{author}{{Wilson Vieira}, A.}, \bibinfo{author}{{Tavares}, G.}, \bibinfo{year}{2003}.
\newblock \bibinfo{title}{Efficient implementation of marching cubes' cases with topological guarantees}.
\newblock \bibinfo{journal}{Journal of Graphics Tools} \bibinfo{volume}{8}, \bibinfo{pages}{1--15}.
\newblock \URLprefix \url{https://doi.org/10.1080/10867651.2003.10487582}, \DOIprefix\doi{10.1080/10867651.2003.10487582}, \href{http://arxiv.org/abs/https://doi.org/10.1080/10867651.2003.10487582}{{\tt arXiv:https://doi.org/10.1080/10867651.2003.10487582}}.
%Type = Misc
\bibitem[{{Louys} et~al.(2017){Louys}, {Tody}, {Dowler}, {Durand}, {Michel}, {Bonnarel}, {Micol} and {IVOA DataModel Working Group}}]{obscore1.1_2017}
\bibinfo{author}{{Louys}, M.}, \bibinfo{author}{{Tody}, D.}, \bibinfo{author}{{Dowler}, P.}, \bibinfo{author}{{Durand}, D.}, \bibinfo{author}{{Michel}, L.}, \bibinfo{author}{{Bonnarel}, F.}, \bibinfo{author}{{Micol}, A.}, \bibinfo{author}{{IVOA DataModel Working Group}}, \bibinfo{year}{2017}.
\newblock \bibinfo{title}{{Observation Data Model Core Components, its Implementation in the Table Access Protocol Version 1.1}}.
\newblock \bibinfo{howpublished}{IVOA Recommendation 09 May 2017}.
\newblock \DOIprefix\doi{10.5479/ADS/bib/2017ivoa.spec.0509L}.
%Type = Misc
\bibitem[{{Mantelet} et~al.(2023){Mantelet}, {Morris}, {Demleitner}, {Dowler}, {Lusted}, {Nieto-Santisteban}, {Ohishi}, {O'Mullane}, {Ortiz}, {Osuna}, {Shirasaki} and {Szalay}}]{adql2023}
\bibinfo{author}{{Mantelet}, G.}, \bibinfo{author}{{Morris}, D.}, \bibinfo{author}{{Demleitner}, M.}, \bibinfo{author}{{Dowler}, P.}, \bibinfo{author}{{Lusted}, J.}, \bibinfo{author}{{Nieto-Santisteban}, M.A.}, \bibinfo{author}{{Ohishi}, M.}, \bibinfo{author}{{O'Mullane}, W.}, \bibinfo{author}{{Ortiz}, I.}, \bibinfo{author}{{Osuna}, P.}, \bibinfo{author}{{Shirasaki}, Y.}, \bibinfo{author}{{Szalay}, A.}, \bibinfo{year}{2023}.
\newblock \bibinfo{title}{{Astronomical Data Query Language Version 2.1}}.
\newblock \bibinfo{howpublished}{IVOA Recommendation 15 December 2023}.
%Type = Inproceedings
\bibitem[{{McMullin} et~al.(2007){McMullin}, {Waters}, {Schiebel}, {Young} and {Golap}}]{CASA2007}
\bibinfo{author}{{McMullin}, J.P.}, \bibinfo{author}{{Waters}, B.}, \bibinfo{author}{{Schiebel}, D.}, \bibinfo{author}{{Young}, W.}, \bibinfo{author}{{Golap}, K.}, \bibinfo{year}{2007}.
\newblock \bibinfo{title}{{CASA Architecture and Applications}}, in: \bibinfo{editor}{{Shaw}, R.A.}, \bibinfo{editor}{{Hill}, F.}, \bibinfo{editor}{{Bell}, D.J.} (Eds.), \bibinfo{booktitle}{Astronomical Data Analysis Software and Systems XVI}, p. \bibinfo{pages}{127}.
%Type = Inproceedings
\bibitem[{{Molinaro} et~al.(2020){Molinaro}, {Allen}, {Genova}, {Schaaff}, {Castro Neves}, {Demleitner}, {Bertocco}, {Morris}, {Bonnarel}, {Voutsinas}, {Boisson} and {Taffoni}}]{IVOA_ESAP2020}
\bibinfo{author}{{Molinaro}, M.}, \bibinfo{author}{{Allen}, M.}, \bibinfo{author}{{Genova}, F.}, \bibinfo{author}{{Schaaff}, A.}, \bibinfo{author}{{Castro Neves}, M.}, \bibinfo{author}{{Demleitner}, M.}, \bibinfo{author}{{Bertocco}, S.}, \bibinfo{author}{{Morris}, D.}, \bibinfo{author}{{Bonnarel}, F.}, \bibinfo{author}{{Voutsinas}, S.}, \bibinfo{author}{{Boisson}, C.}, \bibinfo{author}{{Taffoni}, G.}, \bibinfo{year}{2020}.
\newblock \bibinfo{title}{{The Virtual Observatory ecosystem facing the European Open Science Cloud}}, in: \bibinfo{editor}{{Adler}, D.S.}, \bibinfo{editor}{{Seaman}, R.L.}, \bibinfo{editor}{{Benn}, C.R.} (Eds.), \bibinfo{booktitle}{Observatory Operations: Strategies, Processes, and Systems VIII}, p. \bibinfo{pages}{114491S}.
\newblock \DOIprefix\doi{10.1117/12.2562322}, \href{http://arxiv.org/abs/2103.08334}{{\tt arXiv:2103.08334}}.
%Type = Inproceedings
\bibitem[{{Molinaro} et~al.(2016){Molinaro}, {Butora}, {Bandieramonte}, {Becciani}, {Brescia}, {Cavuoti}, {Costa}, {Di Giorgio}, {Elia}, {Hajnal}, {Gabor}, {Kacsuk}, {Liu}, {Molinari}, {Riccio}, {Schisano}, {Sciacca}, {Smareglia} and {Vitello}}]{vlkb2016}
\bibinfo{author}{{Molinaro}, M.}, \bibinfo{author}{{Butora}, R.}, \bibinfo{author}{{Bandieramonte}, M.}, \bibinfo{author}{{Becciani}, U.}, \bibinfo{author}{{Brescia}, M.}, \bibinfo{author}{{Cavuoti}, S.}, \bibinfo{author}{{Costa}, A.}, \bibinfo{author}{{Di Giorgio}, A.M.}, \bibinfo{author}{{Elia}, D.}, \bibinfo{author}{{Hajnal}, A.}, \bibinfo{author}{{Gabor}, H.}, \bibinfo{author}{{Kacsuk}, P.}, \bibinfo{author}{{Liu}, S.J.}, \bibinfo{author}{{Molinari}, S.}, \bibinfo{author}{{Riccio}, G.}, \bibinfo{author}{{Schisano}, E.}, \bibinfo{author}{{Sciacca}, E.}, \bibinfo{author}{{Smareglia}, R.}, \bibinfo{author}{{Vitello}, F.}, \bibinfo{year}{2016}.
\newblock \bibinfo{title}{{VIALACTEA knowledge base homogenizing access to Milky Way data}}, in: \bibinfo{editor}{{Chiozzi}, G.}, \bibinfo{editor}{{Guzman}, J.C.} (Eds.), \bibinfo{booktitle}{Software and Cyberinfrastructure for Astronomy IV}, p. \bibinfo{pages}{99130H}.
\newblock \DOIprefix\doi{10.1117/12.2231674}, \href{http://arxiv.org/abs/1608.04526}{{\tt arXiv:1608.04526}}.
%Type = Article
\bibitem[{{Namumba} et~al.(2021){Namumba}, {Koribalski}, {J{\'o}zsa}, {Lee-Waddell}, {Jones}, {Carignan}, {Verdes-Montenegro}, {Ianjamasimanana}, {de Blok}, {Cluver}, {Garrido}, {S{\'a}nchez-Exp{\'o}sito}, {Ramaila}, {Thorat}, {Andati}, {Hugo}, {Kleiner}, {Kamphuis}, {Serra}, {Smirnov}, {Maccagni}, {Makhathini}, {Moln{\'a}r}, {Perkins}, {Ramatsoku}, {White} and {Loi}}]{Narumba2021meerkat_debris}
\bibinfo{author}{{Namumba}, B.}, \bibinfo{author}{{Koribalski}, B.S.}, \bibinfo{author}{{J{\'o}zsa}, G.I.G.}, \bibinfo{author}{{Lee-Waddell}, K.}, \bibinfo{author}{{Jones}, M.G.}, \bibinfo{author}{{Carignan}, C.}, \bibinfo{author}{{Verdes-Montenegro}, L.}, \bibinfo{author}{{Ianjamasimanana}, R.}, \bibinfo{author}{{de Blok}, W.J.G.}, \bibinfo{author}{{Cluver}, M.}, \bibinfo{author}{{Garrido}, J.}, \bibinfo{author}{{S{\'a}nchez-Exp{\'o}sito}, S.}, \bibinfo{author}{{Ramaila}, A.J.T.}, \bibinfo{author}{{Thorat}, K.}, \bibinfo{author}{{Andati}, L.A.L.}, \bibinfo{author}{{Hugo}, B.V.}, \bibinfo{author}{{Kleiner}, D.}, \bibinfo{author}{{Kamphuis}, P.}, \bibinfo{author}{{Serra}, P.}, \bibinfo{author}{{Smirnov}, O.M.}, \bibinfo{author}{{Maccagni}, F.M.}, \bibinfo{author}{{Makhathini}, S.}, \bibinfo{author}{{Moln{\'a}r}, D.C.}, \bibinfo{author}{{Perkins}, S.}, \bibinfo{author}{{Ramatsoku}, M.}, \bibinfo{author}{{White}, S.V.}, \bibinfo{author}{{Loi}, F.}, \bibinfo{year}{2021}.
\newblock \bibinfo{title}{{MeerKAT-64 discovers wide-spread tidal debris in the nearby NGC 7232 galaxy group}}.
\newblock \bibinfo{journal}{\mnras} \bibinfo{volume}{505}, \bibinfo{pages}{3795--3809}.
\newblock \DOIprefix\doi{10.1093/mnras/stab1524}, \href{http://arxiv.org/abs/2105.10428}{{\tt arXiv:2105.10428}}.
%Type = Article
\bibitem[{{Pence} et~al.(2010){Pence}, {Chiappetti}, {Page}, {Shaw} and {Stobie}}]{FITS2010}
\bibinfo{author}{{Pence}, W.D.}, \bibinfo{author}{{Chiappetti}, L.}, \bibinfo{author}{{Page}, C.G.}, \bibinfo{author}{{Shaw}, R.A.}, \bibinfo{author}{{Stobie}, E.}, \bibinfo{year}{2010}.
\newblock \bibinfo{title}{{Definition of the Flexible Image Transport System (FITS), version 3.0}}.
\newblock \bibinfo{journal}{\aap} \bibinfo{volume}{524}, \bibinfo{pages}{A42}.
\newblock \DOIprefix\doi{10.1051/0004-6361/201015362}.
%Type = Article
\bibitem[{Punzo et~al.(2017)Punzo, van der Hulst, Roerdink, Fillion-Robin and Yu}]{slicerastro2017}
\bibinfo{author}{Punzo, D.}, \bibinfo{author}{van der Hulst, J.}, \bibinfo{author}{Roerdink, J.}, \bibinfo{author}{Fillion-Robin, J.}, \bibinfo{author}{Yu, L.}, \bibinfo{year}{2017}.
\newblock \bibinfo{title}{Slicerastro: A 3-d interactive visual analytics tool for hi data}.
\newblock \bibinfo{journal}{Astronomy and Computing} \bibinfo{volume}{19}, \bibinfo{pages}{45--59}.
\newblock \URLprefix \url{https://www.sciencedirect.com/science/article/pii/S2213133717300173}, \DOIprefix\doi{https://doi.org/10.1016/j.ascom.2017.03.004}.
%Type = Misc
\bibitem[{{Salgado} et~al.(2024){Salgado}, {Joshi}, {Sánchez-Expósito}, {Parra-Royón}, {Walder}, {Llopis}, {Guo}, {Morris}, {Taffoni}, {Gheller}, {Ouellette}, {Kang}, {Gaudet}, {An}, {Akahori}, {Swinbank}, {Hess}, {Oonk} and {Kok}}]{srcnet0.1_devplan}
\bibinfo{author}{{Salgado}, J.}, \bibinfo{author}{{Joshi}, R.}, \bibinfo{author}{{Sánchez-Expósito}, S.}, \bibinfo{author}{{Parra-Royón}, M.}, \bibinfo{author}{{Walder}, J.}, \bibinfo{author}{{Llopis}, P.}, \bibinfo{author}{{Guo}, S.}, \bibinfo{author}{{Morris}, D.}, \bibinfo{author}{{Taffoni}, G.}, \bibinfo{author}{{Gheller}, C.}, \bibinfo{author}{{Ouellette}, J.}, \bibinfo{author}{{Kang}, H.}, \bibinfo{author}{{Gaudet}, S.}, \bibinfo{author}{{An}, T.}, \bibinfo{author}{{Akahori}, T.}, \bibinfo{author}{{Swinbank}, J.}, \bibinfo{author}{{Hess}, K.M.}, \bibinfo{author}{{Oonk}, R.}, \bibinfo{author}{{Kok}, T.}, \bibinfo{year}{2024}.
\newblock \bibinfo{title}{{SRCNet v0.1 Implementation Plan}}.
\newblock \URLprefix \url{https://confluence.skatelescope.org/display/SNC/SRCNet+Documents}.
%Type = Article
\bibitem[{{S{\'a}nchez} et~al.(2012){S{\'a}nchez}, {Kennicutt}, {Gil de Paz}, {van de Ven}, {V{\'\i}lchez}, {Wisotzki}, {Walcher}, {Mast}, {Aguerri}, {Albiol-P{\'e}rez}, {Alonso-Herrero}, {Alves}, {Bakos}, {Bart{\'a}kov{\'a}}, {Bland-Hawthorn}, {Boselli}, {Bomans}, {Castillo-Morales}, {Cortijo-Ferrero}, {de Lorenzo-C{\'a}ceres}, {Del Olmo}, {Dettmar}, {D{\'\i}az}, {Ellis}, {Falc{\'o}n-Barroso}, {Flores}, {Gallazzi}, {Garc{\'\i}a-Lorenzo}, {Gonz{\'a}lez Delgado}, {Gruel}, {Haines}, {Hao}, {Husemann}, {Igl{\'e}sias-P{\'a}ramo}, {Jahnke}, {Johnson}, {Jungwiert}, {Kalinova}, {Kehrig}, {Kupko}, {L{\'o}pez-S{\'a}nchez}, {Lyubenova}, {Marino}, {M{\'a}rmol-Queralt{\'o}}, {M{\'a}rquez}, {Masegosa}, {Meidt}, {Mendez-Abreu}, {Monreal-Ibero}, {Montijo}, {Mour{\~a}o}, {Palacios-Navarro}, {Papaderos}, {Pasquali}, {Peletier}, {P{\'e}rez}, {P{\'e}rez}, {Quirrenbach}, {Rela{\~n}o}, {Rosales-Ortega}, {Roth}, {Ruiz-Lara}, {S{\'a}nchez-Bl{\'a}zquez}, {Sengupta}, {Singh}, {Stanishev}, {Trager}, {Vazdekis}, {Viironen}, {Wild},
  {Zibetti} and {Ziegler}}]{califa2012}
\bibinfo{author}{{S{\'a}nchez}, S.F.}, \bibinfo{author}{{Kennicutt}, R.C.}, \bibinfo{author}{{Gil de Paz}, A.}, \bibinfo{author}{{van de Ven}, G.}, \bibinfo{author}{{V{\'\i}lchez}, J.M.}, \bibinfo{author}{{Wisotzki}, L.}, \bibinfo{author}{{Walcher}, C.J.}, \bibinfo{author}{{Mast}, D.}, \bibinfo{author}{{Aguerri}, J.A.L.}, \bibinfo{author}{{Albiol-P{\'e}rez}, S.}, \bibinfo{author}{{Alonso-Herrero}, A.}, \bibinfo{author}{{Alves}, J.}, \bibinfo{author}{{Bakos}, J.}, \bibinfo{author}{{Bart{\'a}kov{\'a}}, T.}, \bibinfo{author}{{Bland-Hawthorn}, J.}, \bibinfo{author}{{Boselli}, A.}, \bibinfo{author}{{Bomans}, D.J.}, \bibinfo{author}{{Castillo-Morales}, A.}, \bibinfo{author}{{Cortijo-Ferrero}, C.}, \bibinfo{author}{{de Lorenzo-C{\'a}ceres}, A.}, \bibinfo{author}{{Del Olmo}, A.}, \bibinfo{author}{{Dettmar}, R.J.}, \bibinfo{author}{{D{\'\i}az}, A.}, \bibinfo{author}{{Ellis}, S.}, \bibinfo{author}{{Falc{\'o}n-Barroso}, J.}, \bibinfo{author}{{Flores}, H.}, \bibinfo{author}{{Gallazzi}, A.},
  \bibinfo{author}{{Garc{\'\i}a-Lorenzo}, B.}, \bibinfo{author}{{Gonz{\'a}lez Delgado}, R.}, \bibinfo{author}{{Gruel}, N.}, \bibinfo{author}{{Haines}, T.}, \bibinfo{author}{{Hao}, C.}, \bibinfo{author}{{Husemann}, B.}, \bibinfo{author}{{Igl{\'e}sias-P{\'a}ramo}, J.}, \bibinfo{author}{{Jahnke}, K.}, \bibinfo{author}{{Johnson}, B.}, \bibinfo{author}{{Jungwiert}, B.}, \bibinfo{author}{{Kalinova}, V.}, \bibinfo{author}{{Kehrig}, C.}, \bibinfo{author}{{Kupko}, D.}, \bibinfo{author}{{L{\'o}pez-S{\'a}nchez}, {\'A}.R.}, \bibinfo{author}{{Lyubenova}, M.}, \bibinfo{author}{{Marino}, R.A.}, \bibinfo{author}{{M{\'a}rmol-Queralt{\'o}}, E.}, \bibinfo{author}{{M{\'a}rquez}, I.}, \bibinfo{author}{{Masegosa}, J.}, \bibinfo{author}{{Meidt}, S.}, \bibinfo{author}{{Mendez-Abreu}, J.}, \bibinfo{author}{{Monreal-Ibero}, A.}, \bibinfo{author}{{Montijo}, C.}, \bibinfo{author}{{Mour{\~a}o}, A.M.}, \bibinfo{author}{{Palacios-Navarro}, G.}, \bibinfo{author}{{Papaderos}, P.}, \bibinfo{author}{{Pasquali}, A.},
  \bibinfo{author}{{Peletier}, R.}, \bibinfo{author}{{P{\'e}rez}, E.}, \bibinfo{author}{{P{\'e}rez}, I.}, \bibinfo{author}{{Quirrenbach}, A.}, \bibinfo{author}{{Rela{\~n}o}, M.}, \bibinfo{author}{{Rosales-Ortega}, F.F.}, \bibinfo{author}{{Roth}, M.M.}, \bibinfo{author}{{Ruiz-Lara}, T.}, \bibinfo{author}{{S{\'a}nchez-Bl{\'a}zquez}, P.}, \bibinfo{author}{{Sengupta}, C.}, \bibinfo{author}{{Singh}, R.}, \bibinfo{author}{{Stanishev}, V.}, \bibinfo{author}{{Trager}, S.C.}, \bibinfo{author}{{Vazdekis}, A.}, \bibinfo{author}{{Viironen}, K.}, \bibinfo{author}{{Wild}, V.}, \bibinfo{author}{{Zibetti}, S.}, \bibinfo{author}{{Ziegler}, B.}, \bibinfo{year}{2012}.
\newblock \bibinfo{title}{{CALIFA, the Calar Alto Legacy Integral Field Area survey. I. Survey presentation}}.
\newblock \bibinfo{journal}{\aap} \bibinfo{volume}{538}, \bibinfo{pages}{A8}.
\newblock \DOIprefix\doi{10.1051/0004-6361/201117353}, \href{http://arxiv.org/abs/1111.0962}{{\tt arXiv:1111.0962}}.
%Type = Article
\bibitem[{{Sciacca} et~al.(2015){Sciacca}, {Becciani}, {Costa}, {Vitello}, {Massimino}, {Bandieramonte}, {Krokos}, {Riggi}, {Pistagna} and {Taffoni}}]{visivo2015}
\bibinfo{author}{{Sciacca}, E.}, \bibinfo{author}{{Becciani}, U.}, \bibinfo{author}{{Costa}, A.}, \bibinfo{author}{{Vitello}, F.}, \bibinfo{author}{{Massimino}, P.}, \bibinfo{author}{{Bandieramonte}, M.}, \bibinfo{author}{{Krokos}, M.}, \bibinfo{author}{{Riggi}, S.}, \bibinfo{author}{{Pistagna}, C.}, \bibinfo{author}{{Taffoni}, G.}, \bibinfo{year}{2015}.
\newblock \bibinfo{title}{{An integrated visualization environment for the virtual observatory: Current status and future directions}}.
\newblock \bibinfo{journal}{Astronomy and Computing} \bibinfo{volume}{11}, \bibinfo{pages}{146--154}.
\newblock \DOIprefix\doi{10.1016/j.ascom.2015.01.006}.
%Type = Misc
\bibitem[{{SKA Observatory}(2021)}]{SKA_ConstructionProporsal}
\bibinfo{author}{{SKA Observatory}}, \bibinfo{year}{2021}.
\newblock \bibinfo{title}{{SKA Phase 1 Construction proposal}}.
\newblock \URLprefix \url{https://skao.canto.global/s/M8159?viewIndex=0&column=document&id=bg07p5lsdt2ep2iv9660tpd701}.
%Type = Misc
\bibitem[{{SKA Observatory}(2022)}]{SKA_DeliveryPlan}
\bibinfo{author}{{SKA Observatory}}, \bibinfo{year}{2022}.
\newblock \bibinfo{title}{{SKA Observatory Establishment and Delivery Plan}}.
\newblock \URLprefix \url{https://skao.canto.global/s/M8159?viewIndex=0&column=document&id=bg07p5lsdt2ep2iv9660tpd701}.
%Type = Misc
\bibitem[{{Swinbank} et~al.(2022){Swinbank}, {Bartocco}, {Russo} and {S'{a}nchez-Exp'{o}sito}}]{ESAP2022}
\bibinfo{author}{{Swinbank}, J.D.}, \bibinfo{author}{{Bartocco}, S.}, \bibinfo{author}{{Russo}, S.A.}, \bibinfo{author}{{S'{a}nchez-Exp'{o}sito}, S.}, \bibinfo{year}{2022}.
\newblock \bibinfo{title}{Esap: The escape esfri science analysis platform}.
\newblock \URLprefix \url{https://doi.org/10.5281/zenodo.6044637}, \DOIprefix\doi{10.5281/zenodo.6044637}.
%Type = Article
\bibitem[{{Taylor}(2015)}]{FRELLED2015}
\bibinfo{author}{{Taylor}, R.}, \bibinfo{year}{2015}.
\newblock \bibinfo{title}{{FRELLED: A realtime volumetric data viewer for astronomers}}.
\newblock \bibinfo{journal}{Astronomy and Computing} \bibinfo{volume}{13}, \bibinfo{pages}{67--79}.
\newblock \DOIprefix\doi{10.1016/j.ascom.2015.10.002}, \href{http://arxiv.org/abs/1510.03589}{{\tt arXiv:1510.03589}}.
%Type = Misc
\bibitem[{{Tody} et~al.(2012){Tody}, {Dolensky}, {McDowell}, {Bonnarel}, {Budavari}, {Busko}, {Micol}, {Osuna}, {Salgado}, {Skoda}, {Thompson}, {Valdes} and {Data Access Layer Working Group}}]{ssa1.1}
\bibinfo{author}{{Tody}, D.}, \bibinfo{author}{{Dolensky}, M.}, \bibinfo{author}{{McDowell}, J.}, \bibinfo{author}{{Bonnarel}, F.}, \bibinfo{author}{{Budavari}, T.}, \bibinfo{author}{{Busko}, I.}, \bibinfo{author}{{Micol}, A.}, \bibinfo{author}{{Osuna}, P.}, \bibinfo{author}{{Salgado}, J.}, \bibinfo{author}{{Skoda}, P.}, \bibinfo{author}{{Thompson}, R.}, \bibinfo{author}{{Valdes}, F.}, \bibinfo{author}{{Data Access Layer Working Group}}, \bibinfo{year}{2012}.
\newblock \bibinfo{title}{{Simple Spectral Access Protocol Version 1.1}}.
\newblock \bibinfo{howpublished}{IVOA Recommendation 10 February 2012}.
\newblock \DOIprefix\doi{10.5479/ADS/bib/2012ivoa.spec.0210T}, \href{http://arxiv.org/abs/1203.5725}{{\tt arXiv:1203.5725}}.
%Type = Article
\bibitem[{Vitello et~al.(2018)Vitello, Sciacca, Becciani, Costa, Bandieramonte, Benedettini, Brescia, Butora, Cavuoti, Giorgio, Elia, Liu, Molinari, Molinaro, Riccio, Schisano and Smareglia}]{vialactea2018}
\bibinfo{author}{Vitello, F.}, \bibinfo{author}{Sciacca, E.}, \bibinfo{author}{Becciani, U.}, \bibinfo{author}{Costa, A.}, \bibinfo{author}{Bandieramonte, M.}, \bibinfo{author}{Benedettini, M.}, \bibinfo{author}{Brescia, M.}, \bibinfo{author}{Butora, R.}, \bibinfo{author}{Cavuoti, S.}, \bibinfo{author}{Giorgio, A.M.D.}, \bibinfo{author}{Elia, D.}, \bibinfo{author}{Liu, S.J.}, \bibinfo{author}{Molinari, S.}, \bibinfo{author}{Molinaro, M.}, \bibinfo{author}{Riccio, G.}, \bibinfo{author}{Schisano, E.}, \bibinfo{author}{Smareglia, R.}, \bibinfo{year}{2018}.
\newblock \bibinfo{title}{Vialactea visual analytics tool for star formation studies of the galactic plane}.
\newblock \bibinfo{journal}{Publications of the Astronomical Society of the Pacific} \bibinfo{volume}{130}, \bibinfo{pages}{084503}.
\newblock \URLprefix \url{https://dx.doi.org/10.1088/1538-3873/aac5d2}, \DOIprefix\doi{10.1088/1538-3873/aac5d2}.
%Type = Article
\bibitem[{{Voelsen} et~al.(2021){Voelsen}, {Schachtschneider} and {Brenner}}]{adas2021}
\bibinfo{author}{{Voelsen}, M.}, \bibinfo{author}{{Schachtschneider}, J.}, \bibinfo{author}{{Brenner}, C.}, \bibinfo{year}{2021}.
\newblock \bibinfo{title}{{Classification and Change Detection in Mobile Mapping LiDAR Point Clouds}}.
\newblock \bibinfo{journal}{PFG {\textendash} Journal of Photogrammetry} \bibinfo{volume}{89}, \bibinfo{pages}{195--207}.
\newblock \DOIprefix\doi{10.1007/s41064-021-00148-x}.
%Type = Article
\bibitem[{{Vogt} et~al.(2016){Vogt}, {Owen}, {Verdes-Montenegro} and {Borthakur}}]{Vogt2016}
\bibinfo{author}{{Vogt}, F.P.A.}, \bibinfo{author}{{Owen}, C.I.}, \bibinfo{author}{{Verdes-Montenegro}, L.}, \bibinfo{author}{{Borthakur}, S.}, \bibinfo{year}{2016}.
\newblock \bibinfo{title}{{Advanced Data Visualization in Astrophysics: The X3D Pathway}}.
\newblock \bibinfo{journal}{\apj} \bibinfo{volume}{818}, \bibinfo{pages}{115}.
\newblock \DOIprefix\doi{10.3847/0004-637X/818/2/115}, \href{http://arxiv.org/abs/1510.02796}{{\tt arXiv:1510.02796}}.
%Type = Article
\bibitem[{Vogt et~al.(2017)Vogt, Seitenzahl, Dopita and Ruiter}]{Vogt2017}
\bibinfo{author}{Vogt, F.P.A.}, \bibinfo{author}{Seitenzahl, I.R.}, \bibinfo{author}{Dopita, M.A.}, \bibinfo{author}{Ruiter, A.J.}, \bibinfo{year}{2017}.
\newblock \bibinfo{title}{Linking the x3d pathway to integral field spectrographs: Ysnr 1e 0102.2-7219 in the smc as a case study}.
\newblock \bibinfo{journal}{Publications of the Astronomical Society of the Pacific} \bibinfo{volume}{129}, \bibinfo{pages}{058012}.
\newblock \URLprefix \url{https://dx.doi.org/10.1088/1538-3873/129/975/058012}, \DOIprefix\doi{10.1088/1538-3873/129/975/058012}.
%Type = Article
\bibitem[{{Wilkinson} et~al.(2016){Wilkinson}, {Dumontier}, {Aalbersberg}, {Appleton}, {Axton}, {Baak}, {Blomberg}, {Boiten}, {da Silva Santos}, {Bourne}, {Bouwman}, {Brookes}, {Clark}, {Crosas}, {Dillo}, {Dumon}, {Edmunds}, {Evelo}, {Finkers}, {Gonzalez-Beltran}, {Gray}, {Groth}, {Goble}, {Grethe}, {Heringa}, {'T Hoen}, {Hooft}, {Kuhn}, {Kok}, {Kok}, {Lusher}, {Martone}, {Mons}, {Packer}, {Persson}, {Rocca-Serra}, {Roos}, {van Schaik}, {Sansone}, {Schultes}, {Sengstag}, {Slater}, {Strawn}, {Swertz}, {Thompson}, {van der Lei}, {van Mulligen}, {Velterop}, {Waagmeester}, {Wittenburg}, {Wolstencroft}, {Zhao} and {Mons}}]{FairPrinciples2016}
\bibinfo{author}{{Wilkinson}, M.D.}, \bibinfo{author}{{Dumontier}, M.}, \bibinfo{author}{{Aalbersberg}, I.J.}, \bibinfo{author}{{Appleton}, G.}, \bibinfo{author}{{Axton}, M.}, \bibinfo{author}{{Baak}, A.}, \bibinfo{author}{{Blomberg}, N.}, \bibinfo{author}{{Boiten}, J.W.}, \bibinfo{author}{{da Silva Santos}, L.B.}, \bibinfo{author}{{Bourne}, P.E.}, \bibinfo{author}{{Bouwman}, J.}, \bibinfo{author}{{Brookes}, A.J.}, \bibinfo{author}{{Clark}, T.}, \bibinfo{author}{{Crosas}, M.}, \bibinfo{author}{{Dillo}, I.}, \bibinfo{author}{{Dumon}, O.}, \bibinfo{author}{{Edmunds}, S.}, \bibinfo{author}{{Evelo}, C.T.}, \bibinfo{author}{{Finkers}, R.}, \bibinfo{author}{{Gonzalez-Beltran}, A.}, \bibinfo{author}{{Gray}, A.J.G.}, \bibinfo{author}{{Groth}, P.}, \bibinfo{author}{{Goble}, C.}, \bibinfo{author}{{Grethe}, J.S.}, \bibinfo{author}{{Heringa}, J.}, \bibinfo{author}{{'T Hoen}, P.A.C.}, \bibinfo{author}{{Hooft}, R.}, \bibinfo{author}{{Kuhn}, T.}, \bibinfo{author}{{Kok}, R.}, \bibinfo{author}{{Kok}, J.},
  \bibinfo{author}{{Lusher}, S.J.}, \bibinfo{author}{{Martone}, M.E.}, \bibinfo{author}{{Mons}, A.}, \bibinfo{author}{{Packer}, A.L.}, \bibinfo{author}{{Persson}, B.}, \bibinfo{author}{{Rocca-Serra}, P.}, \bibinfo{author}{{Roos}, M.}, \bibinfo{author}{{van Schaik}, R.}, \bibinfo{author}{{Sansone}, S.A.}, \bibinfo{author}{{Schultes}, E.}, \bibinfo{author}{{Sengstag}, T.}, \bibinfo{author}{{Slater}, T.}, \bibinfo{author}{{Strawn}, G.}, \bibinfo{author}{{Swertz}, M.A.}, \bibinfo{author}{{Thompson}, M.}, \bibinfo{author}{{van der Lei}, J.}, \bibinfo{author}{{van Mulligen}, E.}, \bibinfo{author}{{Velterop}, J.}, \bibinfo{author}{{Waagmeester}, A.}, \bibinfo{author}{{Wittenburg}, P.}, \bibinfo{author}{{Wolstencroft}, K.}, \bibinfo{author}{{Zhao}, J.}, \bibinfo{author}{{Mons}, B.}, \bibinfo{year}{2016}.
\newblock \bibinfo{title}{{The FAIR Guiding Principles for scientific data management and stewardship}}.
\newblock \bibinfo{journal}{Scientific Data} \bibinfo{volume}{3}, \bibinfo{pages}{160018}.
\newblock \DOIprefix\doi{10.1038/sdata.2016.18}.
%Type = Article
\bibitem[{Zou et~al.(2019)Zou, Kiviniemi, Jones and Walsh}]{BIM2019}
\bibinfo{author}{Zou, Y.}, \bibinfo{author}{Kiviniemi, A.}, \bibinfo{author}{Jones, S.}, \bibinfo{author}{Walsh, J.}, \bibinfo{year}{2019}.
\newblock \bibinfo{title}{Risk information management for bridges by integrating risk breakdown structure into 3d/4d bim}.
\newblock \bibinfo{journal}{KSCE Journal of Civil Engineering} \bibinfo{volume}{23}, \bibinfo{pages}{1--14}.
\newblock \DOIprefix\doi{10.1007/s12205-018-1924-3}.

\end{thebibliography}

\end{document}